\documentclass[preprint,showpacs,onecolumn,nofootinbib]{revtex4}
\pdfoutput=1
\usepackage[colorlinks=true,linkcolor=blue,urlcolor=blue,filecolor=black,citecolor=red,pdfstartview=FitV,pdftitle={},pdfsubject={},pdfkeywords={},pdfpagemode=None,bookmarksopen=true]{hyperref}
\usepackage{graphicx}%Include figure files
\usepackage{amsfonts}
\usepackage{amssymb,ulem}
\usepackage{color}%
\usepackage{dcolumn}% Align table columns on decimal pointl
\usepackage{amsmath}
\usepackage{epsfig}
\usepackage{graphics}
\usepackage[active]{srcltx}
\usepackage{epstopdf}
\usepackage{shuffle}
\usepackage[utf8]{inputenc}
\usepackage{bbm}

\newcommand{\be}{\begin{equation}}
\newcommand{\ee}{\end{equation}}
\newcommand{\bea}{\begin{eqnarray}}
\newcommand{\eea}{\end{eqnarray}}
\newcommand{\bean}{\begin{eqnarray*}}
\newcommand{\eean}{\end{eqnarray*}}
\newcommand{\nn}{\nonumber \\}

\def\W #1{\widetilde{#1}}

\def\eref#1{(\ref{#1})}

\def\a{{\alpha}}

\def\b{{\beta}}

\def\eps{\epsilon}

\def\Label#1{\label{#1}%
  \smash{\hbox to0pt{\raise1ex\hbox{\tiny[#1]}\hss}}}

\begin{document}

\baselineskip=0.6 cm
\title{Transmuting off-shell CHY integrals in the double-cover framework}
\author{Kang Zhou}
\email{zhoukang@yzu.edu.cn}
\author{Guo-Jun Zhou}
\email{zhou1750819726@163.com}

\affiliation{Center for Gravitation and Cosmology, College of Physical Science and Technology, Yangzhou University, Yangzhou, 225009, China}

%\date{\today }
\begin{abstract}
\baselineskip=0.6 cm
In this paper, by defining off-shell amplitudes as off-shell CHY integrals, and redefining the longitudinal operator, we demonstrate that the differential operators which link on-shell amplitudes
for a variety of theories together, also link off-shell amplitudes in the similar manner. Based on the algebraic property of the differential operator, we also generalize three relations among color-ordered on-shell amplitudes, including the color-ordered reversed relation, the photon decoupling relation, the Kleiss-Kuijf relation,
to off-shell ones. The off-shell CHY integrals are chosen to be in the double-cover framework, thus,
as a by product, our result also provides a verification
for the double-cover construction.

\end{abstract}


%\pacs{ 04.20.Gz, 04.20.-q, 03.65.-w}
\maketitle

%\tableofcontents

%%%%%%%%%%%%%%%%%%%%%%%%%%%%%%%%%%%%%%
\section{Introduction}
\label{secintro}
%%%%%%%%%%%%%%%%%

The Cachazo-He-Yuan (CHY) formalism reveal the marvelous unity among on-shell tree-level amplitudes of various theories \cite{Cachazo:2013gna,Cachazo:2013hca,Cachazo:2013iea,Cachazo:2014nsa,Cachazo:2014xea}. In the CHY formulae, different
theories correspond to different CHY integrands. Through the so-called dimensional reduction, squeezing,
as well as the generalized dimensional reduction procedures, on-shell CHY integrands for a wide range of theories can be generated from the on-shell CHY integrand for the gravity (GR) theory\footnote{In this paper the
 gravity theory is understood as the Einstein-Hilbert gravity couples to a dilaton and two-forms.}  \cite{Cachazo:2014xea}.

Similar unified web for
on-shell tree-level amplitudes of different theories can be achieved via the differential operators
proposed by Cheung, Shen and Wen.
These differential operators, which act on Lorentz invariant kinematic variables,
transmute the on-shell amplitude of one theory to that of another theory
\cite{Cheung:2017ems}. By applying these operators, amplitudes for a variety of
theories, can be generated from the GR amplitude.
The similarity between two unified webs
implies the underlying relationship between two methods. This relationship has been spelled out in \cite{Zhou:2018wvn,Bollmann:2018edb},
by acting differential operators on CHY integrals for different theories.

Based on the relations among on-shell amplitudes expressed by differential operators, several other relations for
on-shell amplitudes can be derived. First, using the algebraic property of the differential operator, one can derive the generalized
color-ordered reversed relation, the generalized photon decoupling relation, as well as the generalized Kleiss-Kuijf (KK) relation,
which are valid for all color-ordered on-shell amplitudes \cite{Feng:2019cbe}. Secondly, a dual version of the unified web, which reflects the relations
among amplitudes by expanding the amplitude of one theory to amplitudes of another theory, can be established via the differential operators \cite{Feng:2019cbe,Hu:2019qdq,Zhou:2019mbe}. Notice that the expansions
of amplitudes have been studied from various angles in the literature \cite{Stieberger:2016lng,Schlotterer:2016cxa,Chiodaroli:2017ngp,DelDuca:1999rs,Nandan:2016pya,
delaCruz:2016gnm,Fu:2017uzt,Teng:2017tbo,Du:2017kpo,Du:2017gnh}.
Among these angles, the approach used in \cite{Feng:2019cbe,Hu:2019qdq,Zhou:2019mbe} manifests the duality between the differential operators and the coefficients of basis in the expansions.

In this paper, we demonstrate that the differential operators transmute off-shell amplitudes in the manner similar as that for on-shell amplitudes. The off-shell tree-level amplitudes in this paper are defined by off-shell CHY integrals. There are three motivations for considering off-shell amplitudes in the CHY framework. First, the massive external momentum with $k_i^2\neq 0$ can be treated as a special case of the off-shell massless momentum, thus, considering off-shell CHY integrals is an effective way to generalize the CHY formula to massive external states. Secondly, in the double-cover prescription proposed by Gomez,
the evaluation of tree-level amplitudes in the CHY framework can be reduced to lower-point off-shell amplitudes, due to the factorizations realized by the double-cover method \cite{Gomez:2016bmv,Cardona:2016bpi,Bjerrum-Bohr:2018lpz,Gomez:2018cqg,Bjerrum-Bohr:2018jqe,Gomez:2019cik}. This is another important application of off-shell CHY integrals. Thirdly, from the theoretical point of view, it is interesting to broaden the CHY formalism to reproduce off-shell tree-level amplitudes, and for loops to be computed, without a Lagrangian.

The main method used in this paper is similar as
that used in \cite{Zhou:2018wvn,Bollmann:2018edb}, and can be summarized as follows.
The off-shell tree-level amplitude in the CHY formula arise from the
integral over auxiliary coordinates as
\bea
{\cal A}_n=\int d\mu_n\,{\cal I}^{\rm CHY}\,.~~~~\label{CHY-0}
\eea
The measure part $d\mu_n$ is universal, while the CHY integrand ${\cal I}^{\rm CHY}$ depends on the theory under consideration.
All the operators which will be considered in this paper act on Lorentz invariants $\eps_i\cdot \eps_j$ or $\eps_i\cdot k_j$, where $k_i$ and $\epsilon_i$
are the momentum and the polarization vector of the $i^{\rm th}$ leg, respectively. Thus, these operators will not
affect the measure part which is independent of polarization vectors. In other words, differential operators are commutable with the integral
over auxiliary variables.
Thus, transmuting an amplitude is equivalent to transmuting the corresponding CHY integrand. More explicitly,
suppose ${\cal O}$ is a differential operator, it satisfies
\bea
{\cal O}{\cal A}_n={\cal O}\Big(\int d\mu_n\,{\cal I}^{\rm CHY}\Big)=\int d\mu_n\,\Big({\cal O}{\cal I}^{\rm CHY}\Big)\,.
\eea
Thus, if two amplitudes ${\cal A}_\a$ and ${\cal A}_\b$ are related by an operator ${\cal O}$ as ${\cal A}_\a={\cal O}{\cal A}_\b$, analogous relation ${\cal I}^{\rm CHY}_\a={\cal O}\,{\cal I}^{\rm CHY}_\b$ for two integrands
must hold, and vice versa. Consequently, one can derive the unifying relations systematically
by applying differential operators to CHY integrands.

Using the above method, we will show that the unifying relations provided in \cite{Cheung:2017ems}, which link on-shell amplitudes of different theories together, also hold for off-shell amplitudes defined by off-shell CHY integrals.
There are three types of basic operators, the trace operator ${\cal T}[i,j]$, the insertion operator ${\cal I}_{ikj}$, as well as the longitudinal operators ${\cal L}_i$ and ${\cal L}_{ij}$, will be considered in this paper. The definitions of the operators ${\cal T}[i,j]$ and ${\cal I}_{ikj}$
are the same as in \cite{Cheung:2017ems,Zhou:2018wvn,Bollmann:2018edb}, while the definitions of the longitudinal operators ${\cal L}_i$ and ${\cal L}_{ij}$ will be modified in the off-shell case. By applying the combinatory operators constructed by these three basic operators to the GR CHY integral, one can get the CHY integrals for theories including: Einstein-Yang-Mills (EYM) theory, Yang-Mills (YM) theory, Einstein-Maxwell (EM) theory, Einstein-Maxwell theory with photons carry flavors (EMf), Born-Infeld (BI) theory, Yang-Mills-scalar (YMS) theory, special Yang-Mills-scalar (sYMS) theory, bi-adjoint scalar (BAS) theory, non-linear sigma model (NLSM), $\phi^4$ theory, Dirac-Born-Infeld (DBI) theory, extended Dirac-Born-Infeld (exDBI) theory, special Galileon (SG) theory.
Since the unifying relations can be extended to off-shell amplitudes, the relevant relations among color-ordered amplitudes, including the generalized color-ordered reversed relation, the generalized photon decoupling relation, and the generalized KK relation, can also be generalized to the off-shell case.

In \cite{Zhou:2018wvn,Bollmann:2018edb}, CHY integrals are chosen to be in the original single-cover version. In this paper, we choose CHY integrals
in the double-cover formulae \cite{Gomez:2016bmv,Cardona:2016bpi,Bjerrum-Bohr:2018lpz,Gomez:2018cqg,Bjerrum-Bohr:2018jqe,Gomez:2019cik}.
Since on-shell amplitudes can be regarded as the special case of off-shell ones, if the unifying relations hold for off-shell
amplitudes, they also hold for on-shell ones. Thus, our result in this paper provides a verification of the double-cover construction, because we have shown that the unifying relations for on-shell amplitudes in the single-cover formulae, which are already proved in \cite{Zhou:2018wvn,Bollmann:2018edb}, are also correct for off-shell amplitudes in the double-cover framework.

The remainder of this paper is organized as follows. In \S\ref{secreview}, we give a brief introduction of the off-shell CHY formalism, and the double-cover prescription, which are necessary for subsequent discussions. In \S\ref{secbasic}, we study the effects of  three basic operators when applying them to building blocks of CHY integrands.
In \S\ref{secpro} we consider the effects of combinatory operators constructed by these basic operators. The relations among amplitudes, based on the previous preparations, are presented in \S\ref{secunify}. Finally, we end with a brief discussion in \S\ref{secconclu}.

%%%%%%%%%%%%%%%%%%%%%%%%%%%%%%%%%%%%%%
\section{Background}
\label{secreview}
%%%%%%%%%%%%%%%%%

For reader's convenience, in this section we rapidly review the off-shell CHY formalism, and the double-cover prescription.

%%%%%%%%%%%%%%%%%%%%%%%
\subsection{Off-shell CHY formalism}
%%%%%%%%%%%%%%%%%%%%%

The off-shell CHY formalism bears strong similarity with the on-shell one, except the correction elements $\Delta_{ij}$
and $\eta_{ij}$ in scattering equations and matrix elements. In this subsection, we only introduce the off-shell CHY
formalism. The on-shell one can be reproduced by taking $\Delta_{ij}\to 0$, $\eta_{ij}\to 0$.

In the off-shell CHY formula, the tree-level amplitude for $n$ massless particles arises from a multi-dimensional contour integral over
the moduli space of genus zero Riemann surface with $n$ punctures, ${\cal M}_{0,n}$. It can be expressed as
\bea
{\cal A}_n=\int d\mu_n\,{\cal I}_L(\{k_i,\epsilon_i,z_i\}){\cal I}_R(\{k_i,\W\epsilon_i,z_i\})\,,~~~~\label{CHY}
\eea
which possesses the M\"obius ${\rm SL}(2,\mathbb{C})$ invariance. Here $k_i$, $\epsilon_i$ (or $\W\epsilon_i$) and $z_i$ are the momentum, polarization vector, and puncture location for the $i^{\rm th}$
external leg, respectively. The measure is defined as
\bea
d\mu_n\equiv{d^n z\over{\rm vol}\,{\rm SL}(2,\mathbb{C})}{|pqr|\over \prod_{i=1,i\neq pqr}^n\,{\cal E}_i(z)}\,.
\eea
The factor $|pqr|$ is given by $|pqr|\equiv z_{pq}z_{qr}z_{rp}$, where $z_{ij}\equiv z_i-z_j$. The off-shell scattering equations are given as\footnote{In this paper, we choose $2k_i\cdot k_j$ rather than $s_{ij}$ to define the scattering equations. Two choices are un-equivalent for off-shell momenta.} \cite{Naculich:2015zha,Lam:2019mfk,Bjerrum-Bohr:2019nws}
\bea
{\cal E}_i(z)\equiv\sum_{j\in\{1,2,\ldots,n\}\setminus\{i\}}{2k_i\cdot k_j+2\Delta_{ij}\over z_{ij}}=0\,,~~~~\label{SCeqS}
\eea
where
\bea
\Delta_{i,i\pm1}&=&+{1\over2}(k_i^2+k_{i\pm1}^2)\,,\nn
\Delta_{i\mp1,i\pm1}&=&-{1\over2}k_i^2\,,\nn
\Delta_{ij}&=&0~~~~{\rm otherwise}\,.~~~~\label{solu-corr-a}
\eea
These scattering equations yield correct propagators in the Feynman gauge, and satisfy the condition
\bea
\sum_{i=1}^n\,z_i^m{\cal E}_i=0,~~~~m=0,1,2\,,
\eea
which protects the ${\rm SL}(2,\mathbb{C})$ invariance.
The $(n-3)$ independent scattering equations define the map from the space of
kinematic variables to ${\cal M}_{0,n}$, and fully localize the integral on
their solutions.
After fixing the ${\rm SL}(2,\mathbb{C})$ gauge, the measure part is turned to
\bea
d\mu_n\equiv{\big(\prod^n_{j=1,j\neq p,q,r}\,dz_j\big)|pqr|^2\over \prod_{i=1,i\neq p,q,r}^n\,{\cal E}_i(z)}\,.
\eea

The integrand in \eref{CHY} depends on the theory under consideration. For any theory known to have a CHY representation, the corresponding integrand can be factorized into two
parts ${\cal I}_L$ and ${\cal I}_R$, as can be seen in \eref{CHY}. Either of them are weight-$2$ for each puncture coordinate $z_i$
under the ${\rm SL}(2,\mathbb{C})$ transformation. In Table \ref{tab:theories}, we list integrands for
theories which will be considered in this paper\footnote{For theories contain gauge or flavor groups, we only show
the integrands for color-ordered partial amplitudes instead of full ones.} \cite{Cachazo:2014xea}.
\begin{table}[!h]
    \begin{center}
        \begin{tabular}{c|c|c}
            Theory& ${\cal I}_L(\{k_i,\epsilon_i,z_i\})$ & ${\cal I}_R(\{k_i,\W\epsilon_i,z_i\})$ \\
            \hline
           GR & ${\bf Pf}'\Psi$ & ${\bf Pf}'{\Psi}$ \\
            EYM & ${\cal C}_{{\rm Tr}_1}\cdots{\cal C}_{{\rm Tr}_m}{\sum_{\{i,j\}}}'{\cal P}_{\{i,j\}}(n,l,m)$ & ${\bf Pf}'{\Psi}$ \\
            YM & ${\cal C}_n(\sigma)$ & ${\bf Pf}' \Psi$ \\
            EM & ${\bf Pf}'[\Psi]_{n-2m,2m:n-2m}{\bf Pf}[X]_{2m}$ & ${\bf Pf}'{\Psi}$ \\
            EMf & ${\bf Pf}'[\Psi]_{n-2m,2m;n-2m}{\bf Pf}[{\cal X}]_{2m}$
            & ${\bf Pf}'{\Psi}$ \\
            BI & $({\bf Pf}'A)^2$ & ${\bf Pf}' \Psi$ \\
            YMS & ${\cal C}_{{\rm Tr}_1}\cdots{\cal C}_{{\rm Tr}_m}{\sum_{\{i,j\}}}'{\cal P}_{\{i,j\}}(n,l,m)$ & ${\cal C}_n(\sigma)$ \\
            sYMS & ${\bf Pf}'[\Psi]_{n-2m,2m;n-2m}{\bf Pf}[{\cal X}]_{2m}$
            & ${\cal C}_n(\sigma)$ \\
            BAS & ${\cal C}_n(\W\sigma)$ & ${\cal C}_n(\sigma)$ \\
            NLSM & $({\bf Pf}' A)^2$ & ${\cal C}_n(\sigma)$ \\
            $\phi^4$ & ${\bf Pf}' A\,{\bf Pf} [X]_n$ & ${\cal C}_n(\sigma)$ \\
            exDBI & ${\cal C}_{{\rm Tr}_1}\cdots{\cal C}_{{\rm Tr}_m}{\sum_{\{i,j\}}}'{\cal P}_{\{i,j\}}(n,l,m)$
            & $({\bf Pf}' A)^2$ \\
            DBI  & ${\bf Pf}'[\Psi]_{n-2m,2m;n-2m}{\bf Pf}[{\cal X}]_{2m}$ & $({\bf Pf}' A)^2$ \\
            SG & $({\bf Pf}'A)^2$ & $({\bf Pf}' A)^2$ \\
        \end{tabular}
    \end{center}
    \caption{\label{tab:theories}Integrands for various theories}
\end{table}

We now explain each building block in turn. There are five kinds of $n\times n$ matrices, which are defined through
\bea
& &A_{ij} = \begin{cases} \displaystyle {k_{i}\cdot k_j+\Delta_{ij}\over z_{ij}} & i\neq j\,,\\
\displaystyle  ~~~ 0 & i=j\,,\end{cases} \qquad\qquad\qquad\qquad B_{ij} = \begin{cases} \displaystyle {\epsilon_i\cdot\epsilon_j\over z_{ij}} & i\neq j\,,\\
\displaystyle ~~~ 0 & i=j\,,\end{cases} \nn
& &C_{ij} = \begin{cases} \displaystyle {k_i \cdot \epsilon_j+\eta_{ij}\over z_{ij}} &\quad i\neq j\,,\\
\displaystyle -\sum_{l=1,\,l\neq j}^n\hspace{-.5em}C_{li} &\quad i=j\,,\end{cases}
\label{ABCmatrix}
\eea
and
\bea
X_{ij}=\begin{cases} \displaystyle \frac{1}{z_{ij}} & i\neq j\,,\\
\displaystyle ~ ~ 0 & i=j\,,\end{cases} \qquad\qquad\qquad\qquad
{\cal X}_{ij}=\begin{cases} \displaystyle \frac{\delta^{I_i,I_j}}{z_{ij}} & i\neq j\,,\\
\displaystyle ~ ~ 0 & i=j\,,\end{cases}
\eea
Elements $\eta_{ij}$ are given as \cite{Lam:2019mfk}
\bea
\eta_{j\pm1,j}&=&{1\over2}\epsilon_j\cdot k_j\,,\nn
\eta_{ij}&=&0,~~~~{\rm otherwise}\,,~~~~\label{solu-corr-c}
\eea
satisfy
\bea
\sum_{i\neq j,i=1}^n\,\eta_{ij}=\epsilon_j\cdot k_j\,.~~~~\label{eq-corr-c}
\eea
They make $C_{ii}$ to be weight-$2$ under the ${\rm SL}(2,\mathbb{C})$ transformation, therefore keep the ${\rm SL}(2,\mathbb{C})$ invariance
of the whole theory. $\delta^{I_i,I_j}$ is the Kronecker symbol, which forbids the interaction between particles with different flavors $I_i$ and $I_j$.
When the dimension of a matrix need to be clarified, we often denote the $n\times n$ matrix $S$ as $[S]_n$. The $2n\times2n$ antisymmetric matrix $\Psi$ can be constructed from the matrices $A$, $B$ and $C$, in the following form
\bea\label{Psi}
\Psi = \left(
         \begin{array}{c|c}
           ~~A~~ &  ~~C~~ \\
           \hline
           -C^{\rm T} & B \\
         \end{array}
       \right)\,.
\eea
The reduced Pfaffian of $\Psi$ is defined as ${\bf Pf}'\Psi={(-)^{i+j}\over z_{ij}}{\bf Pf}\Psi^{ij}_{ij}$,
where the notation $\Psi^{ij}_{ij}$ means the $i^{\rm th}$ and $j^{\rm th}$ rows and columns in the matrix $\Psi$
have been removed (with $1\leq i<j\leq n$).
Analogous notation holds for ${\bf Pf}'A$.

It is worth to emphasize the definition of Pfaffian, since it is crucial for the work in this paper.
For a $2n\times 2n$ antisymmetric matrix $S$, Pfaffian is defined as
\bea
{\bf Pf}S={1\over 2^n n!}\sum_{\sigma\in S_{2n}} {\bf sgn}(\sigma)\prod_{i=1}^n\,S_{\sigma(2i-1),\sigma(2i)}\,,~~~\label{pfa-1}
\eea
where $S_{2n}$ denotes permutations of $2n$ elements and ${\bf sgn}(\sigma)$ is the signature of $\sigma$.
More explicitly, let $\Pi$ be the set of all partitions of $\{1,2,\cdots, 2n\}$ into pairs without regarding to the order.
An element $\a$ in $\Pi$ can be written as
\bea
\a=\{(i_1,j_1),(i_2,j_2),\cdots,(i_n,j_n) \}\,,
\eea
with $i_k<j_k$ and $i_1<i_2<\cdots<i_n$. Now let
\bea
\pi_{\a} = \left(
         \begin{array}{c}
           ~~~1~~~ 2~~~3~~~4~~\cdots~2n-1~~2n~~ \\
           \,\,i_1~~j_1~~i_2~~j_2~~\cdots~~~i_n~~~~~~j_n \\
         \end{array}
       \right)
\eea
be the corresponding permutation of the partition $\a$. If we define
\bea
S_{\a}={\bf sgn}(\pi_{\a})\,S_{i_1j_1}S_{i_2j_2}\cdots S_{i_nj_n}\,,
\eea
then the Pfaffian of the matrix $S$ is given as
\bea
{\bf Pf}S=\sum_{\a\in\Pi}S_{\a}\,.~~~~~\label{pfa}
\eea
From the definition \eref{pfa}, one can observe that in every term $S_{\a}$
of the Pfaffian, each number in the set $\{1,2,\cdots,2n\}$, which serves as the subscript of the matrix element, will appear once and only once. This simple observation indicates that: {\sl each polarization vector $\epsilon_i$
appears once and only once in each term of the reduced Pfaffian ${\bf Pf}'\Psi$.}
This conclusion is important for latter discussions.

The definition of $\Psi$ can be generalized to the $(2a+b)\times(2a+b)$
matrix $[\Psi]_{a,b:a}$  as
\bea
[\Psi]_{a,b:a}=\left(
         \begin{array}{c|c}
           ~~A_{(a+b)\times (a+b)}~~ &  C_{(a+b)\times a} \\
           \hline
            -C^{\rm T}_{a\times (a+b)} & B_{a\times a} \\
         \end{array}
       \right)\,,~~~~\label{psi-aba}
\eea
here $A$ is a $(a+b)\times (a+b)$ matrix, $C$ is a $(a+b)\times a$ matrix, and $B$ is a $a\times a$ matrix.
The definitions of elements in $A$, $B$ and $C$ are the same as in \eref{ABCmatrix}. The reduced Pfaffian ${\bf Pf}'[\Psi]_{a,b:a}$
can be defined in the same manner.

Starting from the matrix $[\Psi]_{a,b:a}$, the polynomial ${\cal P}_{\{i,j\}}(n,l,m)$ is defined by
\bea
{\cal P}_{\{i,j\}}(n,l,m)&=&
{\bf sgn}(\{i,j\})\,
z_{i_1 j_1}\cdots z_{i_m j_m}\,
{\bf Pf}'[\Psi]_{n-l,i_1,j_1,\ldots,i_m,j_m: n-l}\nn
&=&-{\bf sgn}(\{i,j\}')\,
z_{i_1 j_1}\cdots z_{i_{m-1} j_{m-1}}\,
{\bf Pf}[\Psi]_{n-l,i_1,j_1,\ldots,i_{m-1},j_{m-1}: n-l}\,,~~~~\label{multi-trace-poly}
\eea
where $i_k<j_k\in{\rm Tr}_k$ and ${\rm Tr}_k$ are $m$ sets satisfy\footnote{Each set has at least two elements,
so in general we have $l\geq 2m$.}
\bea
{\rm Tr}_1\cup{\rm Tr}_2\cup\cdots\cup{\rm Tr}_m=\{n-l+1,n-l+2,\cdots,n\}\,.
\eea
In the notation $[\Psi]_{n-l,i_1,j_1,\ldots,i_m,j_m: n-l}$, we explicitly write down $\{i_1,j_1,\ldots,i_m,j_m\}$
instead of $2m$, to emphasize the locations of these $2m$ rows and $2m$ columns in the original matrix $\Psi$.
Two signatures ${\bf sgn}(\{i,j\})$ and ${\bf sgn}(\{i,j\}')$
correspond to partitions $\{(i_1,j_1),\cdots,(i_m,j_m)\}$ and $\{(i_1,j_1),\cdots,(i_{m-1},j_{m-1})\}$
respectively, and one can verify ${\bf sgn}(\{i,j\})={\bf sgn}(\{i,j\}')$.
In the second line of \eref{multi-trace-poly},
the reduced Pfaffian is calculated by removing
rows and columns $i_m$ and $j_m$.
Based on the definition of ${\cal P}_{\{i,j\}}(n,l,m)$ in the second line of \eref{multi-trace-poly},
the polynomial ${\sum_{\{i,j\}}}'{\cal P}_{\{i,j\}}(n,l,m)$ is defined as
\bea
{\sum_{\{i,j\}}}'{\cal P}_{\{i,j\}}(n,l,m)\equiv\sum_{\substack{i_1<j_1\in{\rm Tr}_1\\\cdots\\i_{m-1}<j_{m-1}\in{\rm Tr}_{m-1}}}{\cal P}_{\{i,j\}}(n,l,m) \,.~~~\label{Pfa-Pi}
\eea
where the sum is over all possible choices of pairs in each trace-subset ${\rm Tr}_k$.
In the on-shell case, the polynomial ${\sum_{\{i,j\}}}'{\cal P}_{\{i,j\}}(n,l,m)$ defined above equals to the reduced Pfaffian of the matrix $\Pi$, which is constructed from $\Psi$ via the squeezing procedure \cite{Cachazo:2014xea}. In this paper, we will not use the matrix $\Pi$. The advantage of this choice is, the polynomial ${\sum_{\{i,j\}}}'{\cal P}_{\{i,j\}}(n,l,m)$  defined in
\eref{Pfa-Pi} is manifestly weight-$2$ under the ${\rm SL}(2,\mathbb{C})$ transformation, while the reduced Pfaffian of
$\Pi$ does not have manifest weight.

Finally, the Parke-Taylor factor for the ordering $\sigma$ is given as
\bea
{\cal C}_n(\sigma)={1\over z_{\sigma_1\sigma_2}z_{\sigma_2\sigma_3}\cdots z_{\sigma_{n-1}\sigma_n}z_{\sigma_n\sigma_1}}\,.
\eea
It indicates the color-ordering $(\sigma_1,\sigma_2,\cdots ,\sigma_{n-1},\sigma_n)$ among $n$ external legs.

With ingredients introduced above, off-shell CHY integrands for various theories can be defined as in Table \ref{tab:theories}. Some remarks are in order. First, the ${\rm SL}(2,\mathbb{C})$ symmetry plays the central role in the construction of off-shell CHY integrals. From on-shell integrals
to off-shell ones, all corrections $\Delta_{ij}$ and $\eta_{ij}$ are required by the ${\rm SL}(2,\mathbb{C})$ invariance. It is straightforward to see that both ${\cal I}_L$ and ${\cal I}_R$ defined in Table \ref{tab:theories} are weight-$2$ under the ${\rm SL}(2,\mathbb{C})$
transformation, guarantee the ${\rm SL}(2,\mathbb{C})$ invariance of the whole integral. Secondly, all the reduced Pfaffians appear in off-shell CHY integrands in Table \ref{tab:theories} are independent of the removed rows and columns, similar as those in on-shell integrands. One can follow
the method used in \cite{Cachazo:2013hca}, and use
\bea
\sum_{i=1,i\neq j}^n\,k_i\cdot k_j+\Delta_{ij}=0\,,~~~~
\sum_{i=1,i\neq j}^n\,k_i\cdot \epsilon_j+\eta_{ij}=0\,,
\eea
as well as scattering equations, to prove this.
Thirdly, the gauge invariance no longer exist in general. For example, under the replacement $\epsilon_i\to k_i$,
we do not have ${\bf Pf}'\Psi=0$ anymore, unless we take all $\Delta_{ij}$ and $\eta_{ij}$ to be $0$.
But it is quite natural that the off-shell or massive external states violate the gauge invariance.

%%%%%%%%%%%%%%%%%%%%%%%%%%%%
\subsection{Double-cover prescription}
\label{subsecDC}
%%%%%%%%%%%%%%%%%%%%%%%%%%%%

In next sections, all CHY integrals under consideration are in the double-cover formalism.
In this subsection, we introduce how to reformulate the single-cover CHY integral defined in \eref{CHY} into the double-cover one.

The double-cover prescription of CHY construction is given as a contour integral on $n$-punctured double-covered Rieman spheres \cite{Gomez:2016bmv,Cardona:2016bpi,Bjerrum-Bohr:2018lpz,Gomez:2018cqg,Bjerrum-Bohr:2018jqe,Gomez:2019cik}.
Restricted to the curves
$0=C_i\equiv y^2_i-\sigma^2_i+\Lambda^2$ for $i\in\{1,2,\cdots,n\}$, the pairs $(y_1,\sigma_1),\,(y_2,\sigma_2),\cdots,(y_n,\sigma_n)$ serve as coordinates. Then, all ${1\over z_{ij}}$ in the single-covered formula \eref{CHY} are replaced by
\bea
\tau_{ij}\equiv {1\over2}\Big({y_i+y_j+\sigma_{ij}\over y_i}\Big){1\over\sigma_{ij}}\,.
\eea
Especially, the scattering equations are turned to
\bea
0={\cal E}^\tau_i\equiv\sum_{j\in\{1,2,\cdots,n\}\setminus\{i\}}\,(2k_i\cdot k_j+2\Delta_{ij})\tau_{ij}\,.~~~~\label{SCeqDC}
\eea
Amplitudes in such framework are expressed as the contour integral
\bea
{\cal A}_n=\int d\mu^\Lambda_n {{\cal I}^\tau_L(\sigma_i,y_i,k_i,\epsilon_i){\cal I}^\tau_R(\sigma_i,y_i,k_i,\W\epsilon_i)\over {\cal E}^\tau_m}\,,~~\label{DCamp}
\eea
where the measure $d\mu^\Lambda_n$ is defined through
\bea
d\mu^\Lambda_n\equiv {1\over {\rm vol}\,{\rm GL}(2,\mathbb{C})}{d\Lambda\over \Lambda}\Big(\prod_{i=1}^n\,{y_idy_id\sigma_i\over C_i}\Big){\Delta_{pqr}\over \prod_{j\neq p,q,r,m}{\cal E}^\tau_j}\,,
\eea
with $\Delta_{pqr}\equiv(\tau_{pq}\tau_{qr}\tau_{rp})^{-1}$.
Correspondingly, the contour is determined by poles $\Lambda=0$, $C_i=0$, as well as ${\cal E}^\tau_j=0$ for $j\neq p,q,r,m$.
Eliminating the ${\rm GL}(2,\mathbb{C})$ gauge redundancy turns the measure to be
\bea
d\mu^\Lambda_n\equiv {1\over 2^2}{d\Lambda\over \Lambda}\Big(\prod_{i=1}^n\,{y_idy_i\over C_i}\Big)
\Big(\prod_{j\neq p,q,r,m}\,{d\sigma_j\over {\cal E}^\tau_j}\Big)\Delta_{pqr}\Delta_{pqr|m}\,,
\eea
where
\bea
\Delta_{pqr|m}\equiv\sigma_p\Delta_{qrm}-\sigma_q\Delta_{rmp}+\sigma_r\Delta_{mpq}-\sigma_m\Delta_{pqr}\,.
\eea

Then we consider ${\cal I}^\tau_L(\sigma_i,y_i,k_i,\epsilon_i)$ and ${\cal I}^\tau_R(\sigma_i,y_i,k_i,\W\epsilon_i)$, which are obtained from ${\cal I}_L(z_i,k_i,\epsilon_i)$ and ${\cal I}_R(z_i,k_i,\W\epsilon_i)$ in Table \ref{tab:theories} via the replacement ${1\over z_{ij}}\to\tau_{ij}$. To obtain the double-cover form of the integrand which is more convenient for our consideration, we rewrite $\tau_{ij}$ as
\bea
\tau_{ij}={(y\sigma)_i\over y_i}T_{ij}\equiv{(y\sigma)_i\over y_i}{1\over(y\sigma)_i-(y\sigma)_j}\,,~~~\label{tau-T}
\eea
on the support $C_i=C_j=0$.
The advantage of this reformulation is that $T_{ij}$ carries algebraic properties similar to ${1\over z_{ij}}$,
such as antisymmetry, and
\bea
T_{ik}-T_{jk}={T_{ik}T_{jk}\over T_{ji}}\,.
\eea
This similarity allows us to use a lot of technics for single-cover CHY integrals.
Under the replacement \eref{tau-T}, we have\footnote{Maybe the notations ${\cal I}^T_L(T_{ij},k_i,\epsilon_i)$
and ${\cal I}^T_R(T_{ij},k_i,\W\epsilon_i)$ are more suitable. However, when encounter matrices, notations such as $\Psi^T$, $A^T$ will cause
some ambiguity, since we always use $T$ to denote the transpose of the matrix. Thus we choose the superscript to be $\Lambda$ rather than $T$.}
\bea
& &{\cal I}^\tau_L(\sigma_i,y_i,k_i,\epsilon_i)=\Big(\prod_{i=1}^n{(y\sigma)_i\over y_i}\Big){\cal I}^\Lambda_L(T_{ij},k_i,\epsilon_i)\,,\nn
& &{\cal I}^\tau_R(\sigma_i,y_i,k_i,\W\epsilon_i)=\Big(\prod_{i=1}^n{(y\sigma)_i\over y_i}\Big){\cal I}^\Lambda_R(T_{ij},k_i,\W\epsilon_i)\,.
\eea
For example, the Parke-Taylor factor becomes
\bea
{\cal C}^\tau_n(\sigma)=\Big(\prod_{i=1}^n{(y\sigma)_i\over y_i}\Big){\cal C}^\Lambda_n(\sigma)\equiv\Big(\prod_{i=1}^n{(y\sigma)_i\over y_i}\Big)T_{\sigma_1\sigma_2}T_{\sigma_2\sigma_3}\cdots T_{\sigma_n\sigma_1}\,.~~~~\label{PT-T}
\eea
Thus, start from an integrand in the single-cover version, one can first replace all ${1\over z_{ij}}$ by $T_{ij}$, then times the resulting formula by the factor $\Big(\prod_{i=1}^n\,{(y\sigma)_i\over y_i}\Big)$.
Let us take ${\bf Pf}'\Psi$ as the example, the reduced Pfaffian with new coordinates $y_i$ and $\sigma_i$ is given by
\bea
{\bf Pf}'\Psi^\tau\equiv\Big(\prod_{i=1}^n\,{(y\sigma)_i\over y_i}\Big)(-)^{a+b}T_{ab}{\bf Pf}(\Psi^\Lambda)^{ab}_{ab}\,,
\eea
where the matrix $\Psi^\Lambda$ is obtained from $\Psi$ via the replacement ${1\over z_{ij}}\to T_{ij}$.

%%%%%%%%%%%%%%%%%%%%%%%%%%%%%%%%%%%%%%
\section{Effects of basic operators}
\label{secbasic}
%%%%%%%%%%%%%%%%%

As discussed in \S\ref{secintro}, we are interested in acting differential operators on off-shell CHY integrals in the double-cover version.
To achieve the goal, it is sufficient to apply differential operators to ${\cal I}^\Lambda_L(T_{ij},k_i,\epsilon_i)$ and ${\cal I}^\Lambda_R(T_{ij},k_i,\W\epsilon_i)$, since the operators under consideration will not affect both the measure and the factor ${(y\sigma)_i\over y_i}$.
In this section, we will consider the effects
of applying three basic differential operators to the elementary building-blocks of ${\cal I}^\Lambda_L(T_{ij},k_i,\epsilon_i)$ and ${\cal I}^\Lambda_R(T_{ij},k_i,\W\epsilon_i)$, such as
${\bf Pf}'\Psi^\Lambda$, ${\bf Pf}'[\Psi]_{a,b:a}^\Lambda$, as well as $\sum_{\{i,j\}}'{\cal P}_{\{i,j\}}^\Lambda(n,l,m)$.
Among the operators which will be discussed in this section, the trace and insertion operators are the same as those defined in \cite{Cheung:2017ems},
while the definition of longitudinal operators will be modified.

%%%%%%%%%%%%%%%%%%%%%%%%%%%%%%%%%%%%%%
\subsection{Trace operator}
%%%%%%%%%%%%%%%%%

The trace operator ${\cal T}[i,j]$ is defined as \cite{Cheung:2017ems}
\bea
{\cal T}[i,j]\equiv{\partial\over\partial(\epsilon_i\cdot\epsilon_j)}\,.
\eea
If one applies ${\cal T}[i,j]$ to the reduced Pfaffian  ${\bf Pf}'\Psi^\Lambda$, only terms containing factor $(\epsilon_i\cdot\epsilon_j)$ (i.e.,element $\Psi^\Lambda_{i+n,j+n}$)
provide non-vanishing contributions. Since
$\epsilon_i$ and $\epsilon_j$ appear once and only once in each term
of the reduced Pfaffian,
performing the operator ${\cal T}[i,j]$
is equivalent to the replacement
\bea
\epsilon_i\cdot\epsilon_j\to1\,,~~~~\epsilon_i\cdot V\to 0\,,~~~~\epsilon_j\cdot V\to0\,,
\eea
where $V$ denotes vectors $k_l$ or $\epsilon_{l\neq i,j}$. This manipulation is equivalent to the dimensional
reduction procedure in
\cite{Cachazo:2014xea}. Then, we arrive at a new matrix $\W\Psi^\Lambda$,
satisfies
\bea
{\cal T}_{ij}\,{\bf Pf}'\Psi^\Lambda={\bf Pf}'\W\Psi^\Lambda\,.
\eea
Without lose of generality, we assume $\{i,j\}=\{n-1,n\}$\footnote{This assumption can be realized by moving lows
and columns. Since $(n+i)^{\rm th}$ row and column will be moved simultaneously while moving $i^{\rm th}$ ones,
the possible minus sign will be canceled.}, then the new matrix $\W\Psi^\Lambda$ is given by
\bea
\W\Psi^\Lambda = \left(
         \begin{array}{c|c|c}
           ~~A^\Lambda_{n\times n}~~ &  C^\Lambda_{n\times(n-2)} & 0 \\
           \hline
           -(C^\Lambda)^{\rm T}_{(n-2)\times n} & B^\Lambda_{(n-2)\times(n-2)} & 0\\
           \hline
           0 & 0 & X^\Lambda_{2\times2}\\
         \end{array}
       \right)=\left(
         \begin{array}{c|c}
          [\Psi]^\Lambda_{n-2,2:n-2}& 0\\
           \hline
            0 & [X]^\Lambda_2\\
         \end{array}
       \right)\,,~~~~~\label{result-M}
\eea
where the matrices $[\Psi]^\Lambda_{n-2,2:n-2}$ and $[X]^\Lambda_2$ are obtained from $[\Psi]_{n-2,2:n-2}$ and $[X]_2$
through the replacement ${1\over z_{ij}}\to T_{ij}$.
The reduced Pfaffian of the matrix $\W\Psi^\Lambda$ can be calculated straightforwardly as
\bea
{\bf Pf}'\W\Psi^\Lambda={\bf Pf}'[\Psi]^\Lambda_{n-2,2;n-2}{\bf Pf}[X]^\Lambda_2\,.~~~~\label{result1}
\eea
Thus, we find
\bea
{\cal T}[i,j]\,{\bf Pf}'\Psi^\Lambda&=&{\bf Pf}'[\Psi]^\Lambda_{n-2,2;n-2}{\bf Pf}[X]^\Lambda_2\,.~~~~\label{result-trace-1}
\eea

Applying the similar procedure to the matrix $[\Psi]^\Lambda_{n-2,2:n-2}$ gives
\bea
{\cal T}[i,j]{\bf Pf}'[\Psi]^\Lambda_{n-2,2:n-2}={\bf Pf}'[\Psi]^\Lambda_{a-4,4:a-4}{\bf P
f}[X]^\Lambda_2\,.
\eea
By repeating the manipulation, we find the
multiple action of trace operators gives the following recursive pattern
\bea
{\cal T}[i_1,j_1]{\cal T}[i_2,j_2]\,{\bf Pf}'\Psi^\Lambda&=&{\bf Pf}'[\Psi]^\Lambda_{n-4,4;n-4}{\bf Pf}[X_1]^\Lambda_2{\bf Pf}[X_2]^\Lambda_2\,,\nn
& &\cdots\nn
{\cal T}[i_1,j_1]{\cal T}[i_2,j_2]\cdots{\cal T}[i_m,j_m]\,{\bf Pf}'\Psi^\Lambda&=&{\bf Pf}'[\Psi]^\Lambda_{n-2m,2m;n-2m}{\bf Pf}[X_1]^\Lambda_2{\bf Pf}[X_2]^\Lambda_2
\cdots{\bf Pf}[X_m]^\Lambda_2\nn
&=&(-)^m({T_{i_1j_1}}{T_{j_1i_1}})({T_{i_2j_2}}{T_{j_2i_2}})\cdots({T_{i_mj_m}}{T_{j_mi_m}}){\cal P}_{\{i,j\}}^\Lambda(n,2m,m)\,,~~~~\label{poly}
\eea
where the polynomial ${\cal P}_{\{i,j\}}^\Lambda(n,l,m)$ is obtained from ${\cal P}_{\{i,j\}}(n,l,m)$ in \eref{multi-trace-poly} via the replacement ${1\over z_{ij}}\to T_{ij}$, and we have arranged elements as
\bea
[X_k]^\Lambda_2=\left(
         \begin{array}{c|c}
         0& {T_{i_kj_k}}\\
           \hline
           {T_{j_ki_k}} & 0\\
         \end{array}
       \right)\,.
\eea
We want to point out that the factor $(T_{ij}T_{ji})$ appear in \eref{poly} is the simplest Parke-Taylor factor
${\cal C}^\Lambda_2$ defined in \eref{PT-T}, indicates the simplest color-ordering $(i,j)$. More general color-orderings can be generated from it by inserting other elements.

%%%%%%%%%%%%%%%%%%%%%%%%%%%%%%%%%%%%%%
\subsection{Insertion operator}
%%%%%%%%%%%%%%%%%

The insertion operator is defined by \cite{Cheung:2017ems}
\bea
{\cal I}_{ikj}\equiv{\partial\over\partial(k_i\cdot\epsilon_k)}-{\partial\over\partial(k_j\cdot\epsilon_k)}\,.
\eea
In this subsection, we discuss the effect of acting this operator on the
polynomial ${\sum_{\{i,j\}}}'{\cal P}^\Lambda_{\{i,j\}}(n,l,m)$.
As will be seen immediately, the most important effect is
replacing $T_{ij}$
in the Parke-Taylor factor by $T_{ik}T_{kj}$.
In other words, this operator
transmutes the color-ordering
$(\cdots,i,j,\cdots)$ to $(\cdots,i,k,j,\cdots)$.
To show this, one need to assume $i^{\rm th}$ and $j^{\rm th}$ legs belong to the same trace subset, i.e., $i,j\in{\rm Tr}_k$. For simplicity, we assume $i,j\in{\rm Tr}_m$, and
take the expansion \eref{multi-trace-poly} where rows and columns $i_m,j_m\in{\rm Tr}_m$
have been removed (with replacing ${1\over z_{ij}}$ by $T_{ij}$). Since in the off-shell case the reduced Pfaffian
is also independent of the choice of removed rows and columns, as pointed out in the previous section, assuming $i$ and $j$ belong to any other ${\rm Tr}_k$ will
not change the conclusion, although the calculation will be more complicate..

Let us consider the polynomial ${\bf Pf}[\Psi]^\Lambda_{n-l,i_1,j_1,\ldots,i_{m-1},j_{m-1}: n-l}$, which is given as
\bea
{\bf Pf}[\Psi]^\Lambda_{n-l,i_1,j_1,\ldots,i_{m-1},j_{m-1}: n-l}=\sum_{\a\in\Pi}{\bf sgn}(\pi_{\a})[\Psi]^\Lambda_{a_1b_1}[\Psi]^\Lambda_{a_2b_2}\cdots[\Psi]^\Lambda_{a_{(n'+m')}b_{(n'+m')}}\,,~~~~\label{pfaffian-insertion}
\eea
where the definition of Pfaffian in \eref{pfa} has been used. The element $[\Psi]^\Lambda_{a_ib_i}$ is at the $a_i^{\rm th}$ row and $b_i^{\rm th}$ column in the matrix $[\Psi]^\Lambda_{n-l,i_1,j_1,\ldots,i_{m-1},j_{m-1}: n-l}$, and we have defined $n'=n-l$, $m'=m-1$.
Since we have chosen $i,j\in{\rm Tr}_m$, it is straightforward to see that $k_i\cdot\epsilon_k$ appears only in $C^\Lambda_{kk}$. Thus, when applying ${\partial\over\partial(k_i\epsilon_k)}$ to \eref{pfaffian-insertion}, only terms containing the element $[\Psi]^\Lambda_{k,n'+2m'+k}$ (see the formula \eref{psi-aba}) can survive. For
such a term, the remaining part after eliminating $[\Psi]^\Lambda_{k,n'+2m'+k}$ corresponds to a partition of the
the  set $\{1,2,\cdots,2(n'+m')\}\setminus\{k,n'+2m'+k\}$, which has the length $2(n'+m'-1)$. Such a term appears in the  ${\bf Pf}[\Psi]^\Lambda_{n-l-1,i_1,j_1,\ldots,i_{m-1},j_{m-1}: n-l-1}$,
weighted by a new signature ${\bf sgn}(\pi_{\W\a})$, where  the
new matrix $[\Psi]^\Lambda_{n-l-1,i_1,j_1,\ldots,i_{m-1},j_{m-1}: n-l-1}$ is obtained from the original one $[\Psi]^\Lambda_{n-l,i_1,j_1,\ldots,i_{m-1},j_{m-1}: n-l}$ by removing $k^{\rm th}$ and $(n'+2m'+k)^{\rm th}$ rows and columns,
and ${\bf sgn}(\pi_{\W\a})$ corresponds to the partition of the
length-$2(n'+m'-1)$ set. Comparing two special partitions, where one corresponds to the original
matrix, another one corresponds to the new matrix,
\bea
\a&=&\{(a_1,b_1),(a_2,b_2),\cdots,(k,n'+2m'+k),\cdots,(a_{(n'+m')},b_{(n'+m')})\},\nn
\W\a&=&\{(a_1,b_1),(a_2,b_2),\cdots,(a_{(n'+m'-1)},b_{(n'+m'-1)})\}\,,
\eea
one can find  ${\bf sgn}(\pi_{\a})=(-)^{n'-1}{\bf sgn}(\pi_{\W\a})$.
Thus, summing all contributions together gives
\bea
{\partial\over\partial(k_i\cdot\epsilon_k)}{\bf Pf}[\Psi]^\Lambda_{n-l,i_1,j_1,\ldots,i_{m-1},j_{m-1}: n-l}=(-)^{n-l}{T_{ik}}{\bf Pf}[\Psi]^\Lambda_{n-l-1,i_1,j_1,\ldots,i_{m-1},j_{m-1}: n-l-1}\,.
\eea

Applying the above result to \eref{multi-trace-poly}, we get
\bea
{\cal T}_{ikj}\Big({\sum_{\{i,j\}}}'{\cal P}^\Lambda_{\{i,j\}}(n,l,m)\Big)&=&(-)^{n-l}\Big({T_{ik}}-{ T_{jk}}\Big)\Big({\sum_{\{i,j\}}}'{\cal P}^\Lambda_{\{i,j\}}(n,l+1,m)\Big)\nn
&=&(-)^{n-l}{T_{ik}T_{kj}\over T_{ij}}\Big({\sum_{\{i,j\}}}'{\cal P}^\Lambda_{\{i,j\}}(n,l+1,m)\Big)\,,~~~~\label{result-insertion}
\eea
where the definition
\bea
T_{ij}\equiv{1\over (y\sigma)_i}-{1\over (y\sigma)_j}
\eea
has been used.
The factor ${T_{ik}T_{kj}\over T_{ij}}$ in the result \eref{result-insertion} indicates that the insertion operator transmutes $T_{ij}$ in the Parke-Taylor factor to $T_{ik}T_{kj}$, thus inserts the $k^{\rm th}$ external leg between $i^{\rm th}$ and $j^{\rm th}$ legs in the color-ordering.

%%%%%%%%%%%%%%%%%%%%%%%%%%%%%%%%%%%%%%
\subsection{Longitudinal operator}
%%%%%%%%%%%%%%%%%

For on-shell amplitudes, the longitudinal operators are defined via \cite{Cheung:2017ems}
\bea
{\cal L}'_i\equiv\sum_{j\neq i}\,(k_i\cdot k_j){\partial\over\partial(k_j\cdot\epsilon_i)}\,,
\eea
and
\bea
{\cal L}'_{ij}\equiv-(k_i\cdot k_j){\partial\over \partial(\epsilon_i\cdot\epsilon_j)}\,.
\eea
For the off-shell case, they should be modified to
\bea
{\cal L}_i\equiv\sum_{j\neq i}(k_i\cdot k_j+\Delta_{ij}){\partial\over\partial(k_j\cdot\epsilon_i)}\,,
\eea
and
\bea
{\cal L}_{ij}\equiv-(k_i\cdot k_j+\Delta_{ij}){\partial\over \partial(\epsilon_i\cdot\epsilon_j)}\,.
\eea
The reason for the above modifications will be seen in the next section. We now discuss the effects of acting them on the
reduced Pfaffian ${\bf Pf}'[\Psi]^\Lambda_{a,b:a}$.

We first consider the operator ${\cal L}_{ij}$. It turns $(\epsilon_i\cdot\epsilon_j)$ to
$(k_i\cdot k_j+\Delta_{ij})$, and annihilates all other $(\epsilon_i\cdot V)$, $(\epsilon_j\cdot V)$. Using the observation that $\epsilon_i$
and $\epsilon_j$ appear once and only once respectively, one can conclude that ${\cal L}_{ij}$ transmutes
the reduced Pfaffian of the matrix $[\Psi]^\Lambda_{a,b:a}$ as follows
\bea
{\cal L}_{ij}\,{\bf Pf}'\left(
         \begin{array}{c|c}
           ~~A^\Lambda_{(a+b)\times (a+b)}~~ &  C^\Lambda_{(a+b)\times a}  \\
           \hline
           -(C^\Lambda)^{\rm T}_{a\times (a+b)} & B^\Lambda_{a\times a}\\
         \end{array}
       \right)
       \Rightarrow
{\bf Pf}'\left(
         \begin{array}{c|c|c}
           ~~A^\Lambda_{(a+b)\times (a+b)}~~ &  C^\Lambda_{(a+b)\times (a-2)} & 0 \\
           \hline
           -(C^\Lambda)^{\rm T}_{(a-2)\times (a+b)} & B^\Lambda_{(a-2)\times(a-2)} & 0\\
           \hline
           0 & 0 & -A^\Lambda_{2\times2}\\
         \end{array}
       \right)\,.~~~~\label{long1}
\eea

Then, we consider the operator ${\cal L}_i$. It is straightforward to see
\bea
{\partial \eta_{ji}\over\partial (k_j\cdot \epsilon_i)}=0\,,
\eea
thus the operator ${\cal L}_i$ transmutes every $(k_j\cdot\epsilon_i+\eta_{ji})$ to $(k_j\cdot k_i+\Delta_{ij})$.
Under such replacement, the diagonal elements of the matrix $C^\Lambda$
are transmuted to
\bea
C^\Lambda_{ii}\to-\sum_{l=1,\,l\neq i}^n\,(k_l \cdot k_i +\Delta_{lj})T_{li}\,,
\eea
and vanishes due to scattering equations
\bea
0={\cal E}^\tau_i\equiv\sum_{j=1,j\neq i}^n\,(2k_i\cdot k_j+2\Delta_{ij})\tau_{ij}={(y\sigma)_i\over y_i}\sum_{j=1,j\neq i}^n\,(2k_i\cdot k_j+2\Delta_{ij})T_{ij}\,.
\eea
Thus, the operator ${\cal L}_i$ has the following effect
\bea
{\cal L}_i\,{\bf Pf}'\left(
         \begin{array}{c|c}
           ~~A^\Lambda_{(a+b)\times (a+b)}~~ &  C^\Lambda_{(a+b)\times a}  \\
           \hline
           -(C^\Lambda)^{\rm T}_{a\times (a+b)} & B^\Lambda_{a\times a}\\
         \end{array}
       \right)
       \Rightarrow
{\bf Pf}'\left(
         \begin{array}{c|c|c}
           ~~A^\Lambda_{(a+b)\times (a+b)}~~ &  C^\Lambda_{(a+b)\times (a-2)} & A^\Lambda_{(a+b)\times2} \\
           \hline
           -(C^\Lambda)^{\rm T}_{(a-2)\times (a+b)} & B^\Lambda_{(a-2)\times(a-2)} & 0\\
           \hline
           A^\Lambda_{2\times (a+b)} & 0 & 0\\
         \end{array}
       \right)\,.~~~~\label{long2}
\eea
The results \eref{long1} and \eref{long2} are crucial for generating the ingredient $({\bf Pf}'A)^2$.

%%%%%%%%%%%%%%%%%%%%%%%%%%%%%%%%%%%%%%
\section{Effects of combinatory operators}
\label{secpro}
%%%%%%%%%%%%%%%%%

The combinatory operators are constructed by three types of basic operators. In this section, by using the results obtained in the previous section, we will consider three kinds of combinatory
operators, especially their action on the reduced Pfaffian ${\bf
Pf}'\Psi^\Lambda$, which is the fundamental building-block for the GR integrand.

%%%%%%%%%%%%%%%%%%%%%%%%%%%%%%%%%%%
\subsection{Operator ${\cal T}[\a]$}
%%%%%%%%%%%%%%%%%%%%%%%%%%%%%%%%%%%

The general trace operator ${\cal T}[\a]$ for an ordered length-$m$ set
$\a=\{\a_1,\a_2,\cdots,\a_m\}$ is  defined as\footnote{We adopt the
convention in \cite{Cheung:2017ems} that the product of two
operators ${\cal O}_1\cdot{\cal O}_2$ acts on an amplitude as
$({\cal O}_1\cdot{\cal O}_2){\cal A} ={\cal O}_2{\cal O}_1{\cal A}$,
i.e., the operator ${\cal O}_1$ is performed at first, and ${\cal
O}_2$ secondly.} \cite{Cheung:2017ems}
\bea
{\cal T}[\a]\equiv{\cal T}[\a_1,\a_m]\cdot\prod_{i=2}^{m-1}{\cal I}_{\a_{i-1}\a_i\a_m}\,.~~~~\label{defin-trace}
\eea
As discussed before, when acting on ${\bf Pf}'\Psi^\Lambda$, the operator ${\cal T}[\a_1,\a_m]$ creates the Parke-Taylor factor $T_{\a_1\a_m}T_{\a_m\a_1}$.
On the other hand, the insertion operator ${\cal I}_{\a_{i-1}\a_i\a_m}$ transmutes $T_{\a_{i-1}\a_m}$ to $T_{\a_{i-1}\a_i}T_{\a_i\a_m}$.
Thus, we can expect that the operator ${\cal T}[\a]$ will create the Parke-Taylor factor $T_{\a_1\a_2}\cdots T_{\a_{m-1}\a_m}T_{\a_m\a_1}$, which indicates
the color-ordering $(\a_1,\a_2,\cdots,\a_m)$.

To verify this expectation, let us compute the result of applying the operator ${\cal T}[\a]$ to ${\bf Pf}'\Psi^\Lambda$.
At the first step, performing ${\cal T}[\a_1,\a_m]$ gives
\bea
{\cal T}[\a_1,\a_m]\,{\bf Pf}'\Psi^\Lambda&=&T_{\a_1\a_m}{\bf Pf}'[\Psi]^\Lambda_{n-2,\a_1,\a_m:n-2}\nn
&=&{T_{\a_1\a_m}T_{\a_m\a_1}}{\bf Pf}[\Psi]^\Lambda_{n-2:n-2}\,,~~~~\label{begin}
\eea
where the result \eref{result-trace-1} have been used.
Then one can act ${\cal I}_{\a_1\a_2\a_m}$ on the resulting object, and use \eref{result-insertion} to get
\bea
{\cal I}_{\a_1\a_2\a_m}{\cal T}[\a_1,\a_m]\,{\bf Pf}'\Psi^\Lambda&=&(-)^{n-2}{T_{\a_1\a_m}T_{\a_m\a_1}}{T_{\a_1\a_2}T_{\a_2\a_m}\over T_{\a_1\a_m}}{\bf Pf}[\Psi]^\Lambda_{n-3:n-3}\nn
&=&(-)^{n-2}{T_{\a_1\a_2}T_{\a_2\a_m}T_{\a_m\a_1}}{\bf Pf}[\Psi]^\Lambda_{n-3:n-3}\,.
\eea
Similar manipulation gives
\bea
{\cal I}_{\a_2\a_3\a_m}{\cal I}_{\a_1\a_2\a_m}{\cal T}[\a_1,\a_m]\,{\bf Pf}'\Psi^\Lambda&=&(-)^{n-2}(-)^{n-3}{{T_{\a_1\a_2}T_{\a_2\a_m}T_{\a_m\a_1}}
T_{\a_2\a_3}T_{\a_3\a_m}\over {T_{\a_2\a_m}}}{\bf Pf}[\Psi]^\Lambda_{n-4:n-4}\nn
&=&(-)^{(n-2)+(n-3)}{T_{\a_1\a_2}T_{\a_2\a_3}T_{\a_3\a_m}T_{\a_m\a_1}}{\bf Pf}[\Psi]^\Lambda_{n-4:n-4}\,,
\eea
and the recursive pattern can be observed.
Repeating the above procedure, one will arrive at
\bea {\cal T}[\a]\,{\bf Pf}'\Psi^\Lambda&=&(-)^{(2n-m-1)(m-2)\over2}{{T_{\a_1\a_2}T_{\a_2\a_3}\cdots T_{\a_{m-1}\a_m}T_{\a_m\a_1}}}{\bf
Pf}[\Psi]^\Lambda_{n-m:n-m}\nn &=&(-)^{{(2n-m-1)(m-2)\over2}+1}{\cal C}^\Lambda_\a{\sum_{\{i,j\}}}'{\cal
P}^\Lambda_{\{i,j\}}(n,m,1)\,.~~~\label{single-trace} \eea
Especially, when $m=n$, we have
\bea
{\cal T}[\a_1,\a_2,\cdots,\a_n]\,{\bf Pf}'\Psi^\Lambda={(-)^{(n-1)(n-2)\over 2}{ T_{\a_1\a_2}T_{\a_2\a_3}\cdots T_{\a_{n-1}\a_n}T_{\a_n\a_1}}}
=(-)^{(n-1)(n-2)\over 2}{\cal C}^\Lambda_n\,.~~~~\label{rela-2}
\eea

The procedure of generating the color-ordering $(\a_1,\cdots,\a_m)$ can be understood as follows. At the first step, two reference points $\alpha_1$
and $\alpha_m$ are created by the operator ${\cal T}[\alpha_1,\alpha_m]$. Then, other legs in the set $\alpha$ are inserted between $\alpha_1$ and $\alpha_m$,
via insertion operators. This interpretation allows the trace operator ${\cal T}[\alpha]$ to have a variety of equivalent expressions,
since different legs can be inserted by a variety of equivalent ways.
As an example, let us consider the color-ordering $(1,2,3,4,5)$. To generate it, the firs step can be creating two reference points $1$
and $5$ via the trace operator ${\cal T}[1,5]$. Then, one can insert other legs between $1$ and $5$ by the following order:
\begin{itemize}
\item inserts $2$ between $1$ and $5$,
\item inserts $3$ between $2$ and $5$,
\item inserts $4$ between $3$ and $5$.
\end{itemize}
This order yields the operator
\bea
{\cal T}[1,2,3,4,5]={\cal T}[1,5]\cdot{\cal I}_{125}\cdot{\cal I}_{235}\cdot{\cal I}_{345}\,,
\eea
coincide with the initial definition \eref{defin-trace}. However, one can also choose other equivalent orders, for instance:
\begin{itemize}
\item inserts $3$ between $1$ and $5$,
\item inserts $2$ between $1$ and $3$,
\item inserts $4$ between $3$ and $5$,
\end{itemize}
and arrive at the operator
\bea
{\cal T}'[1,2,3,4,5]={\cal T}[1,5]\cdot{\cal I}_{135}\cdot{\cal I}_{123}\cdot{\cal I}_{345}\,,
\eea
When applying to ${\bf Pf}'\Psi^\Lambda$, two operator ${\cal T}[1,2,3,4,5]$ and ${\cal T}'[1,2,3,4,5]$ are equivalent to each other, as can be verified by the technic from \eref{begin} to \eref{single-trace}.

Furthermore, based on the cyclic symmetry of the color-ordering, one can chose arbitrary two points $\alpha_a$ and $\alpha_b$ as
reference points, then insert $\alpha_{a+1},\cdots,\alpha_{b-1}$ between $\alpha_a$ and $\alpha_b$, and insert $\alpha_{b+1},\cdots,\alpha_{a-1}$
between $\alpha_b$ and $\alpha_a$. For example, one can also choose the trace operator to be
\bea
{\cal T}[\a]\equiv{\cal T}[\a_2,\a_{m-1}]\cdot{\cal I}_{\a_{m-1}\a_m\a_2}\cdot{\cal I}_{\a_m\a_1\a_2} \cdot\Big(\prod_{i=3}^{m-2}{\cal I}_{\a_{i-1}\a_i\a_{m-1}}\Big)\,.
\eea

The previous result in the current subsection can be generalized to multi-trace cases ${\cal
T}[\a_1]\cdot{\cal T}[\a_2]\cdots$, with the constraint
$[\a_i]\cap[\a_j]=\emptyset$ for arbitrary $i$ and $j$. Let us consider, for example,
\bea
{\cal T}[\a]\cdot{\cal T}[\b]&=&\Big({\cal T}[\a_1,\a_m]\cdot\prod_{i=2}^{m-1}{\cal I}_{\a_{i-1}\a_i\a_m}\Big)
\cdot\Big({\cal T}[\b_1,\b_l]\cdot\prod_{i=2}^{l-1}{\cal I}_{\b_{i-1}\b_i\b_l}\Big)\nn
&=&{\cal T}[\a_1,\a_m]\cdot{\cal T}[\b_1,\b_l]\cdot\Big(\prod_{i=2}^{m-1}{\cal I}_{\a_{i-1}\a_i\a_m}\Big)
\cdot\Big(\prod_{i=2}^{l-1}{\cal I}_{\b_{i-1}\b_i\b_l}\Big)\,.
\eea
From the first line to the second line, the commutativity of the operators has been used.
First, we apply the operators ${\cal T}[\a_1,\a_m]$ and ${\cal T}[\b_1,\b_l]$ to ${\bf Pf}'\Psi^\Lambda$. Using \eref{poly}, we obtain
\bea
{\cal T}[\b_1,\b_l]{\cal T}[\a_1,\a_m]\,{\bf Pf}'\Psi^\Lambda
&=&\Big({-T_{\b_1\b_l}T_{\b_l\b_1}}\Big)
\Big({-T_{\a_1\a_m}T_{\a_m\a_1}}\Big){\sum_{\{i,j\}}}'{\cal P}^\Lambda_{\{i,j\}}(n,4,2)\,,~~~~\label{step1}
\eea
where
\bea
{\sum_{\{i,j\}}}'{\cal P}^\Lambda_{\{i,j\}}(n,4,2)={\cal P}^\Lambda_{\{i,j\}}(n,4,2)={-1\over T_{\b_1\b_l}}{\bf Pf}[\Psi]^\Lambda_{n-4,\b_1,\b_l:n-4}
={-1\over T_{\a_1\a_m}}{\bf Pf}[\Psi]^\Lambda_{n-4,\a_1,\a_m:n-4}\,.
\eea
Secondly, we use \eref{result-insertion} to get
\bea
& &{\cal I}_{\a_{m-2}\a_{m-1}\a_m}\cdots{\cal I}_{\a_{2}\a_{3}\a_m}{\cal I}_{\a_{1}\a_{2}\a_m}\Big({-T_{\a_1\a_m}T_{\a_m\a_1}}\Big){\sum_{\{i,j\}}}'{\cal P}^\Lambda_{\{i,j\}}(n,4,2)\nn
&=&(-)^{n-4+1}{\cal I}_{\a_{m-2}\a_{m-1}\a_m}\cdots{\cal I}_{\a_{2}\a_{3}\a_m}\Big({T_{\a_1\a_2}T_{\a_2\a_m}T_{\a_m\a_1}}\Big){\sum_{\{i,j\}}}'{\cal P}^\Lambda_{\{i,j\}}(n,5,2)\nn
& &~~~~~~~~~~~~~\cdots\nn
&=&(-)^{{(2n-5-m)(m-2)\over2}+1}{\cal C}^\Lambda_{\a}{\sum_{\{i,j\}}}'{\cal P}^\Lambda_{\{i,j\}}(n,2+m,2)\,.~~~~\label{step2}
\eea
Thirdly, we use \eref{result-insertion} again to obtain
\bea
& &{\cal I}_{\b_{l-2}\a_{l-1}\a_l}\cdots{\cal I}_{\b_{2}\b_{3}\b_l}{\cal I}_{\b_{1}\b_{2}\b_l}\Big({-T_{\b_1\b_l}T_{\b_l\b_1}}\Big){\sum_{\{i,j\}}}'{\cal P}^\Lambda_{\{i,j\}}(n,2+m,2)\nn
&=&(-)^{{(2n-2m-l-1)(l-2)\over2}+1}{\cal C}^\Lambda_{\b}{\sum_{\{i,j\}}}'{\cal P}^\Lambda_{\{i,j\}}(n,l+m,2)\,.~~~~\label{step3}
\eea
Combining \eref{step1}, \eref{step2} and \eref{step3} together, we get
\bea
{\cal T}[\a]\cdot{\cal T}[\b]\,{\bf Pf}'\Psi^\Lambda=(-)^{{(2n-l-m-3)(l+m-4)\over2}+2}{\cal C}^\Lambda_{\a}{\cal C}^\Lambda_{\b}{\sum_{\{i,j\}}}'{\cal P}^\Lambda_{\{i,j\}}(n,l+m,2)\,.
\eea
The most general formula is given by
\bea
{\cal T}[\a_1]\cdot{\cal T}[\a_2]\cdots{\cal T}[\a_k]\,{\bf Pf}'\Psi^\Lambda=(-)^{{(2n-2k-\sum|\a_i|+1)(\sum|\a_i|-2k)\over2}+k}\Big(\prod{\cal C}^\Lambda_{\a_i}\Big){\sum_{\{i,j\}}}'{\cal P}^\Lambda_{\{i,j\}}(n,\sum |\a_i|,k)\,,~~~~\label{rela-3}
\eea
where $|\a_i|$ denotes the length of the set $\a_i$. It can be obtained recursively, by applying the extremely similar technic.

%%%%%%%%%%%%%%%%%%%%%%%%%%%%%%%%%%%%%%
\subsection{Operators ${\cal T}[a,b]\cdot{\cal L}$ and ${\cal T}[a,b]\cdot{\cal \W L}$}
%%%%%%%%%%%%%%%%%

The operator ${\cal L}$ is defined through longitudinal operators as \cite{Cheung:2017ems}
\bea {\cal L}\equiv\prod_i{\cal L}_i={\cal \W L}+\cdots\,,~~~~{\rm with}~{\cal \W L}\equiv
\sum_{\rho\in {\rm pair}} \prod_{i_k,j_k\in\rho}{\cal
L}_{i_kj_k}.~~~\label{LT} \eea
Here the set of pairs $\{(i_1,j_1),(i_2,j_2),\cdots,(i_m,j_m)\}$ is
a partition of $I$ with conditions $i_1<i_2<...<i_m$ and
$i_t<j_t,~\forall t$.
At the algebraic level, two operators ${\cal L}$ and ${\cal \W L}$ are not equivalent to each other.
However, for on-shell integrands, if we apply the combinatory operators ${\cal
T}[a,b]\cdot{\cal L}$ and ${\cal
T}[a,b]\cdot{\cal \W L}$ to ${\bf Pf}'\Psi^\Lambda$ with even number of external legs, and let subscripts of ${\cal L}_i$ and ${\cal L}_{ij}$
run through all nodes in $\{1,2,\cdots,n\}\setminus\{a,b\}$, two operators have the same effect \cite{Cheung:2017ems,Zhou:2018wvn,Bollmann:2018edb}, providing
\bea
{\cal T}[a,b]\cdot{\cal L}\,{\bf Pf}'\Psi^\Lambda={\cal
T}[a,b]\cdot{\cal \W L}\,{\bf Pf}'\Psi^\Lambda\doteq{\bf Pf}'A^\Lambda)^2\,,~~~~\label{eff-LT}
\eea
up to an overall sign. We now show that the relation \eref{eff-LT} also holds for off-shell integrands.

We first consider the effect of the manipulation ${\cal T}[a,b]\cdot{\cal \W L}\,{\bf Pf}'\Psi^\Lambda$. Acting ${\cal T}[a,b]$ on ${\bf Pf}'\Psi^\Lambda$ gives
\eref{result-trace-1}, which is the reduced Pfaffian of the matrix \eref{result-M}. Using the previous result \eref{long1}, it is straightforward to see ${\cal \W L}$ transmutes the matrix
\eref{result-M} to
\bea
\Psi'^\Lambda = \left(
         \begin{array}{c|c|c}
           ~~A^\Lambda_{n\times n}~~ &  0 & 0 \\
           \hline
          0 & -A^\Lambda_{(n-2)\times(n-2)} & 0\\
           \hline
           0 & 0 & X^\Lambda_{2\times2}\\
         \end{array}
       \right)\,,
\eea
The Pfaffian of the matrix
\bea
\left(
         \begin{array}{c|c}

           -A^\Lambda_{(n-2)\times(n-2)} & 0\\
           \hline
           0 & X^\Lambda_{2\times2}\\
         \end{array}
       \right)
\eea
is just $-{\bf Pf}'(-A^\Lambda)=(-)^{{n\over2}}{\bf Pf}'A^\Lambda$, thus
\bea
{\cal T}[a,b]\cdot{\cal \W L}\,{\bf Pf}'\Psi^\Lambda
={\bf Pf}'\Psi'^\Lambda=(-)^{{n\over2}}\Big({\bf Pf}'A^\Lambda\Big)^2\,.
\eea

Then, we consider the effect of acting ${\cal L}$
on ${\cal T}[a,b]\,{\bf Pf}'\Psi^\Lambda$.
Using \eref{long2} we know the operator ${\cal L}$ transmutes
the matrix \eref{result-M} to
\bea
\Psi''^\Lambda= \left(
         \begin{array}{c|c|c}
           ~~A^\Lambda_{n\times n}~~ &  A^\Lambda_{n\times(n-2)} & 0 \\
           \hline
          A^\Lambda_{(n-2)\times n} & 0 & 0\\
           \hline
           0 & 0 & X^\Lambda_{2\times2}\\
         \end{array}
       \right)\,,
\eea
thus the reduced Pfaffian ${\bf Pf}'\Psi^\Lambda$ is turned to
\bea
{\bf Pf}'\Psi''^\Lambda= {\bf Pf}'\W A^\Lambda\,{\bf Pf}[X]^\Lambda_2\,,~~~\label{pf''}
\eea
where
\bea
\W A^\Lambda\equiv \left(
         \begin{array}{c|c}
           ~~A^\Lambda_{n\times n}~~ &  A^\Lambda_{n\times(n-2)} \\
           \hline
          A^\Lambda_{(n-2)\times n} & 0 \\
         \end{array}
       \right)\,.~~~\label{WA}
\eea
For simplicity, we choose $a^{\rm th}$
and $b^{\rm th}$ rows  and columns to be removed when evaluating the reduced Pfaffian of $\W A^\Lambda$. Then we need to compute
\bea
{\bf Pf} \left(
         \begin{array}{c|c}
           ~~(A^\Lambda_{n\times n})^{ab}_{ab}~~ &  (A^\Lambda_{n\times n})^{ab}_{ab} \\
           \hline
          (A^\Lambda_{n\times n})^{ab}_{ab} & 0 \\
         \end{array}
       \right)\,.~~~~\label{pfa-AA}
\eea
Using the definition of Pfaffian \eref{pfa}, one can find that the non-vanishing contributions for \eref{pfa-AA} come from rows $i\in\{1,\cdots,n-2\}$ and columns $j\in\{n-1,\cdots,2n-4\}$, which give rise to the determinate of the matrix $(A^\Lambda)^{ab}_{ab}$.
Thus the reduced Pfaffian of $\W A^\Lambda$
can be obtained as
\bea
{\bf Pf}'\W A^\Lambda=(-)^{{(n-2)(n-3)\over2}+1}T_{ab}{\bf det}(A^\Lambda)^{ab}_{ab}=(-)^{{n\over2}}{1\over T_{ab}}\Big({\bf Pf}'A^\Lambda\Big)^2\,,
\eea
where we have used $(-)^{(n-2)(n-3)\over2}=(-)^{n-2\over2}$, due to the fact $n$ is even. Putting it back to \eref{pf''}, we obtain
\bea
{\cal T}[a,b]\cdot{\cal L}\,{\bf Pf}'\Psi^\Lambda
={\bf Pf}'\Psi''^\Lambda=(-)^{{n\over2}}\Big({\bf Pf}'A^\Lambda\Big)^2\,.
\eea
Above calculations show that
\bea {\cal T}[a,b]\cdot{\cal L}\,{\bf Pf}'\Psi^\Lambda={\cal T}[a,b]\cdot{\cal \W L}\,{\bf Pf}'\Psi^\Lambda =(-)^{{n\over2}}\Big({\bf
Pf}'A^\Lambda\Big)^2\,.~~~~\label{rela-4}
\eea
It is worth to notice that this result is independent of the choice of $a$ and $b$.

Although the relation \eref{eff-LT} for on-shell amplitudes also holds for off-shell ones, we must emphasize that
the definitions of operators ${\cal L}$ and ${\cal \W L}$ are different for the on-shell and off-shell cases, since
for the off-shell case, corrections $\Delta_{ij}$ are introduced when defining longitudinal operators ${\cal L}_i$ and ${\cal L}_{ij}$.
We have shown that, with the redefined ${\cal L}_i$ and ${\cal L}_{ij}$, one can transmute ${\bf Pf}'\Psi^\Lambda$
to $({\bf Pf}'A^\Lambda)^2$, which serves as a building block for BI, DBI, exDBI, NLSM and SG integrands, therefore have meaningful interpretation.
If we insist the original definition, the resulting final object can not be interpreted physically.

%%%%%%%%%%%%%%%%%%%%%%%%%%%%%%%%%%%%%%
\subsection{Operators ${\cal T}_{X_{2m}}$ and ${\cal T}_{{\cal X}_{2m}}$ }
%%%%%%%%%%%%%%%%%

In this subsection, we consider the combinatory operators ${\cal T}_{X_{2m}}$ and ${\cal T}_{{\cal X}_{2m}}$,
which generate ${\bf Pf}[X]^\Lambda_{2m}$ and ${\bf Pf}[{\cal X}]^\Lambda_{2m}$ from ${\bf Pf}'\Psi^\Lambda$, respectively.

For a given length-$2m$ set $I$, the operator ${\cal T}_{X_{2m}}$ is defined as
\bea
{\cal T}_{X_{2m}}\equiv\sum_{\rho\in {\rm pair}}\prod_{i_k,j_k\in\rho}{\cal T}[i_k,j_k]\,.
~~~\label{4.28}\eea
The notation $\sum_{\rho\in {\rm pair}}\prod_{i_k,j_k\in\rho}$ is explained after \eref{LT}. Using the result in \eref{poly}, and the
definition of Pfaffian \eref{pfa}, one can find that the operator ${\cal T}_{X_{2m}}$
transmutes $\Psi^\Lambda$ to a new matrix
\bea
\W\Psi^{\ast\Lambda} = \left(
         \begin{array}{c|c|c}
           ~~A^\Lambda_{n\times n}~~ &  C^\Lambda_{n\times(n-2m)} & 0 \\
           \hline
           -(C^\Lambda)^{\rm T}_{(n-2m)\times n} & B^\Lambda_{(n-2m)\times(n2-m)} & 0\\
           \hline
           0 & 0 & X^\Lambda_{2m\times2m}\\
         \end{array}
       \right)\,,~~~~~\label{result-M'}
\eea
so that
\bea
{\cal T}_{X_{2m}}\,{\bf Pf}'\Psi^\Lambda={\bf Pf}'\W\Psi^{\ast\Lambda}={\bf Pf}'[\Psi]^\Lambda_{n-2m,2m:n-2m}{\bf Pf}[X]^\Lambda_{2m}\,,~~~~\label{rela-5}
\eea
therefore provides the building block ${\bf Pf}[X]^\Lambda_{2m}$.

The operator ${\cal T}_{{\cal X}_{2m}}$ is defined in a similar form as
\bea
{\cal T}_{{\cal X}_{2m}}\equiv\sum_{\rho\in {\rm pair}}\prod_{i_k,j_k\in\rho}\delta^{I_{i_k},I_{j_k}}{\cal T}[i_k,j_k]\,,
~~~\label{4.31}\eea
where $\delta^{I_{i_k},I_{j_k}}$ forbids the interaction between particles with different flavors. These $\delta^{I_{i_k},I_{j_k}}$ turn the matrix $[X]^\Lambda_{2m}$ to $[{\cal X}]^\Lambda_{2m}$.
Thus, we have
\bea
{\cal T}_{{\cal X}_{2m}}\,{\bf Pf}'\Psi^\Lambda={\bf Pf}'[\Psi]^\Lambda_{n-2m,2m:n-2m}{\bf Pf}[{\cal X}]^\Lambda_{2m}\,.~~~~\label{rela-6}
\eea
which gives the building block ${\bf Pf}[{\cal X}]^\Lambda_{2m}$.

%%%%%%%%%%%%%%%%%%%%%%%%%%%%%%%%%%%%%%
\section{Relations among amplitudes}
\label{secunify}
%%%%%%%%%%%%%%%%%

With preparations in previous sections, we are ready to exhibit
relations among different off-shell amplitudes.
We will re-establish
the unified web in
\cite{Cheung:2017ems} for off-shell amplitudes. Then, we will claim that three important relations among color-ordered on-shell amplitudes, including
the color-ordered reversed relation, the photon decoupling relation, as well as the KK relation, can be generalized
to the off-shell case.

%%%%%%%%%%%%%%%%%%%%%%%%%%%%%%%%%%%%%%
\subsection{Transmuting amplitudes by differential operators}
%%%%%%%%%%%%%%%%%

In this subsection, we will show the combinatory differential operators discussed in \S\ref{secpro} transmute the GR amplitude to amplitudes of a variety of other theories.
It is quite natural to take the GR amplitude whose external states carry highest spins as the starting point, since
all operators decrease the spins of external legs. The GR integrand in the single-cover formula is shown in the
first line of Table \ref{tab:theories}. To achieve the double-cover expression, we can get
${\cal I}^\Lambda_L$ and ${\cal I}^\Lambda_R$ from ${\cal I}_L$ and
${\cal I}_R$, via the replacement ${1\over z_{ij}}\to T_{ij}$, then time both of them by the factor $\Big(\prod_{i=1}^n\,{(y\sigma)_i\over y_i}\Big)$. Thus, we have
\bea
{\cal A}_{\rm GR}=\int\,d\mu^\Lambda_n\,\Big(\prod_{i=1}^n\,{(y\sigma)_i\over y_i}\Big)^2{\bf Pf}'\Psi^\Lambda{\bf Pf}'\W\Psi^\Lambda\,.
\eea

We will consider the effect of applying differential operators to the above GR CHY integral. As introduced in \S\ref{secintro}, the
key point is, differential operators are commutable with the integral
over auxiliary coordinates, thus transmuting an
amplitude is equivalent to transmuting the corresponding CHY integrand.
Further more, since differential operators do not affect the factor ${(y\sigma)_i\over y_i}$,
transmuting ${\cal I}_L^\tau$ or ${\cal I}_R^\tau$ is equivalent to transmuting ${\cal I}_L^\Lambda$ or ${\cal I}_R^\Lambda$.
In summary, we have
\bea
{\cal O}{\cal A}&=&{\cal O}\int\,d\mu^\Lambda_n\,\Big(\prod_{i=1}^n\,{(y\sigma)_i\over y_i}\Big)^2{\cal I}^\Lambda_L{\cal I}^\Lambda_R\nn
&=&\int\,d\mu^\Lambda_n\,\Big(\prod_{i=1}^n\,{(y\sigma)_i\over y_i}\Big)^2\Big({\cal O}\,{\cal I}^\Lambda_L{\cal I}^\Lambda_R\Big)\,.
\eea

In the GR integrand, two parts ${\cal I}^\Lambda_L$ and
${\cal I}^\Lambda_R$ depend on two independent sets of polarization vectors
$\{\epsilon_i\}$ and $\{\W\epsilon_i\}$, respectively.
It means we can define two independent sets of differential operators, through two sets of polarization vectors.
The operators defined via $\{\epsilon_i\}$ only act on ${\cal I}^\Lambda_L$, while the operators defined via $\{\W\epsilon_i\}$ only act on ${\cal I}^\Lambda_R$.
This property protects the manifest double copy structure of CHY integrands.
Without lose of generality, we can restrict
the effect of operators on the ${\cal I}^\Lambda_L$ part, by defining them via $\{\epsilon_i\}$.

Performing operators on the ${\cal I}^\Lambda_L$ part and
using \eref{poly}, \eref{rela-2}, \eref{rela-3}, \eref{rela-4},
\eref{rela-5}, as well as \eref{rela-6}, after comparing with the middle column of Table
\ref{tab:theories}, we get following relations:
\bea
{\cal A}_{{\rm EYM}}&=&{\cal T}[{\rm Tr}_1]\cdots{\cal T}[{\rm Tr}_m]\,{\cal A}_{{\rm GR}}\,,\nn
{\cal A}_{{\rm YM}}&=&{\cal T}[i_1\cdots i_n]\,{\cal A}_{{\rm GR}}\,,\nn
{\cal A}_{{\rm EM}}&=&{\cal T}_{X_{2m}}\,{\cal A}_{{\rm GR}}\,,\nn
{\cal A}_{{\rm EMf}}&=&{\cal T}_{{\cal X}_{2m}}\,{\cal A}_{{\rm GR}}\,,\nn
{\cal A}_{{\rm BI}}&=&{\cal T}[a,b]\cdot{\cal L}\,{\cal A}_{{\rm GR}}\,,~~\label{A-rela-1}
\eea
up to an overall sign.

Similarly, applying operators to the pure YM integrand,
we obtain relations:
\bea
{\cal A}_{{\rm YMS}}&=&{\cal T}[{\rm Tr}_1]\cdots{\cal T}[{\rm Tr}_m]\,{\cal A}_{{\rm YM}}\,,\nn
{\cal A}_{{\rm sYMS}}&=&{\cal T}_{{\cal X}_{2m}}\,{\cal A}_{{\rm YM}}\,,\nn
{\cal A}_{{\rm BAS}}&=&{\cal T}[i_1\cdots i_n]\,{\cal A}_{{\rm YM}}\,,\nn
{\cal A}_{{\rm NLSM}}&=&{\cal T}[a,b]\cdot{\cal L}\,{\cal A}_{{\rm YM}}\,,\nn
{\cal A}_{\phi^4}&=&{\cal T}_{X_{n}}\,{\cal A}_{{\rm YM}}\,,~~\label{A-rela-2}
\eea
up to an overall sign. Notice that the amplitude of $\phi^4$ theory is generated
via a special ${\cal T}_{X_{2m}}$ with $2m=n$.

Applying operators to the BI integrand, we get relations:
\bea
{\cal A}_{{\rm exDBI}}&=&{\cal T}[{\rm Tr}_1]\cdots{\cal T}[{\rm Tr}_m]\,{\cal A}_{{\rm BI}}\,,\nn
{\cal A}_{{\rm DBI}}&=&{\cal T}_{{\cal X}_{2m}}\,{\cal A}_{{\rm BI}}\,,\nn
{\cal A}_{{\rm NLSM}}&=&{\cal T}[i_1\cdots i_n]\,{\cal A}_{{\rm BI}}\,,\nn
{\cal A}_{{\rm SG}}&=&{\cal T}[a,b]\cdot{\cal L}\,{\cal A}_{{\rm BI}}\,,~~\label{A-rela-3}
\eea
up to an overall sign.

Relations presented in \eref{A-rela-1}, \eref{A-rela-2} and \eref{A-rela-3} can be organized as
\bea
{\cal A}_{{\rm other}}={\cal O}^{\epsilon}\cdot{\cal O}^{\W\epsilon}\,{\cal A}^{\epsilon,\W\epsilon}_{{\rm G}}\,,~~~~\label{LR}
\eea
where ${\cal O}^{\epsilon}$ and ${\cal O}^{\W\epsilon}$ denote operators which are defined through $\{\epsilon_i\}$ and $\{\W\epsilon_i\}$,
respectively. The corresponding operators ${\cal O}^\epsilon$ and ${\cal O}^{\W\epsilon}$ for different theories are listed in
Table \ref{tab:unifying}.
\begin{table}[!h]
    \begin{center}
        \begin{tabular}{c|c|c}
            Amplitude& ${\cal O}^\epsilon$ & ${\cal O}^{\W\epsilon}$ \\
            \hline
            ${\cal A}_{{\rm EYM}}^{\epsilon,\W\epsilon}$ & ${\cal T}^{\epsilon}[{\rm Tr}_1]\cdots{\cal T}^{\epsilon}[{\rm Tr}_m]$
            & $\mathbb{I}$ \\
            ${\cal A}_{{\rm YM}}^{\W\epsilon}$&  ${\cal T}^{\epsilon}[i_1,\cdots, i_n]$ & $\mathbb{I}$\\
            ${\cal A}_{{\rm EM}}^{\epsilon,\W\epsilon}$ & ${\cal T}^{\epsilon}_{X_{2m}}$ & $\mathbb{I}$\\
            ${\cal A}_{{\rm EMf}}^{\epsilon,\W\epsilon}$&  ${\cal T}^{\epsilon}_{{\cal X}_{2m}}$ & $\mathbb{I}$ \\
            ${\cal A}_{{\rm BI}}^{\W\epsilon}$ & ${\cal L}^{\epsilon}\cdot{\cal T}^{\epsilon}[ab]$ & $\mathbb{I}$ \\
            ${\cal A}_{{\rm YMS}}^{\W\epsilon}$ & ${\cal T}^{\epsilon}[i_1,\cdots, i_n]$ & ${\cal T}^{\W\epsilon}[{\rm Tr}_1]\cdots{\cal T}^{\W\epsilon}[{\rm Tr}_m]$ \\
            ${\cal A}_{{\rm sYMS}}^{\W\epsilon}$ &  ${\cal T}^{\epsilon}[i_1,\cdots, i_n]$ & ${\cal T}^{\W\epsilon}_{{\cal X}_{2m}}$ \\
            ${\cal A}_{{\rm BAS}}$ &  ${\cal T}^{\epsilon}[i_1,\cdots, i_n]$ & ${\cal T}^{\W\epsilon}[i_1',\cdots, i_n']$ \\
            ${\cal A}_{{\rm NLSM}}$ & ${\cal T}^{\epsilon}[i_1,\cdots, i_n]$ & ${\cal L}^{\W\epsilon}\cdot{\cal T}^{\W\epsilon}_{a'b'}$ \\
            ${\cal A}_{\phi^4}$ &  ${\cal T}^{\epsilon}[i_1,\cdots, i_n]$ & ${\cal T}^{\W\epsilon}_{X_n}$ \\
            ${\cal A}_{{\rm exDBI}}^{\W\epsilon}$ &  ${\cal L}^{\epsilon}\cdot{\cal T}^{\epsilon}[ab]$ & ${\cal T}^{\W\epsilon}[{\rm Tr}_1]\cdots{\cal T}^{\W\epsilon}[{\rm Tr}_m]$ \\
            ${\cal A}_{{\rm DBI}}^{\W\epsilon}$ & ${\cal L}^{\epsilon}\cdot{\cal T}^{\epsilon}[ab]$ & ${\cal T}^{\W\epsilon}_{{\cal X}_{2m}}$ \\
            ${\cal A}_{{\rm SG}}$ &  ${\cal L}^{\epsilon}\cdot{\cal T}^{\epsilon}[ab]$ & ${\cal L}^{\W\epsilon}\cdot{\cal T}^{\W\epsilon}[a'b']$ \\
        \end{tabular}
    \end{center}
    \caption{\label{tab:unifying}Unifying relations}
\end{table}
In this table, $\mathbb{I}$ denotes the identical operator. As pointed out before,
since the manifest double copy structure of the CHY integrands, ${\cal O}^{\epsilon}$ and ${\cal O}^{\W\epsilon}$ act on two pieces independently at the integrand-level.

From relations mentioned in Table \ref{tab:unifying}, various other relations can be extracted. For example, substituting
\bea
{\cal A}^{\epsilon,\W\epsilon}_{\rm EYM}={\cal T}^\epsilon[{\rm Tr}_1]\cdots{\cal T}^\epsilon[{\rm Tr}_m]{\cal A}^{\epsilon,\W\epsilon}_{\rm GR}
\eea
into
\bea
{\cal A}^{\epsilon}_{\rm YMS}=\Big({\cal T}^\epsilon[{\rm Tr}_1]\cdots{\cal T}^\epsilon[{\rm Tr}_m]\Big)\cdot{\cal T}^{\W\epsilon}[i_1,\cdots,i_n]{\cal A}^{\epsilon,\W\epsilon}_{\rm GR}\,,
\eea
we get
\bea
{\cal A}^{\epsilon}_{\rm YMS}={\cal T}^{\W\epsilon}[i_1,\cdots,i_n]{\cal A}^{\epsilon,\W\epsilon}_{\rm EYM}\,.
\eea

Differential operators connect not only amplitudes of different theories, but also amplitudes of
the same theory.
For example, using the relation \eref{result-insertion} one can get
\bea
{\cal A}^{\epsilon,\W\epsilon}_{\rm EYM}(\{i_3^h,\cdots, i_m^h\}||j_1^g,i_1^g,i_2^g,j_2^g,\cdots, j_n^g)
&=&{\cal T}^\epsilon_{i_1i_2j_2}\,{\cal T}^\epsilon_{j_1i_1j_2}\,{\cal A}^{\epsilon,\W\epsilon}_{\rm EYM}(\{i_1^h,\cdots, i_m^h\}||j_1^g,\cdots, j_n^g)\,.
\eea
The notation $(\{i_1^h,\cdots, i_m^h\}||j_1^g,\cdots, j_n^g)$ means there is no color-ordering among elements in the set $\{i_1^h,\cdots, i_m^h\}$,
while elements $(j_1^g,\cdots, j_n^g)$ are color-ordered.
The insertion operators turn gravitons $i_1^h$ and $i_2^h$ to gluons, and insert them between gluons $j_1^g$ and $j_2^g$
in the color ordering $(j_1^g,\cdots,j_n^g)$. Consequently, insertion operators ${\cal T}^\epsilon_{i_1i_2j_2}\,,{\cal T}^\epsilon_{j_1i_1j_2}$
transmutes the EYM amplitude ${\cal A}^{\epsilon,\W\epsilon}_{\rm EYM}(\{i_1^h,\cdots, i_m^h\}||j_1^g,\cdots, j_n^g)$ to the EYM amplitude ${\cal A}^{\epsilon,\W\epsilon}_{\rm EYM}(\{i_3^h,\cdots, i_m^h\}||j_1^g,i_1^g,i_2^g,j_2^g,\cdots, j_n^g)$.

%%%%%%%%%%%%%%%%%%%%%%%%%%%%%%%%%%%%%%%
\subsection{Three generalized relations for color-ordered amplitudes}
%%%%%%%%%%%%%%%%%%%%%%%%%%%%%%%%%%%%%%%

In this subsection, we demonstrate that three generalized relations for on-shell color-ordered amplitudes, also hold for off-shell amplitudes.
From Table \ref{tab:unifying}, one can see that each color-ordered amplitude can be generated by applying the general trace operator ${\cal T}[\a]$,
formally expressed as
\bea
{\cal A}(i_1,\cdots,i_n)={\cal T}[i_1,\cdots,i_n]{\cal A}'\,.
\eea
Thus, we will derived three relations by using the algebraic property
of the general trace operator.

The first relation is the generalized color-ordered reversed relation for off-shell color-ordered amplitudes:
\bea
{\cal A}(1,2,\cdots,n)=(-)^n{\cal A}(n,n-1,\cdots,1)\,.~~~~\label{color-reverse}
\eea
The color-ordering $(1,2,\cdots,n;\cdots)$ can be created by the general trace operator
\bea
{\cal T}[1,2,\cdots,n]={\cal T}[1,n]\cdot{\cal I}_{12n}\cdot{\cal I}_{23n}\cdots{\cal I}_{(n-2)(n-1)n}\,.
\eea
Using ${\cal T}[1,n]={\cal T}[n,1]$ and ${\cal I}_{ikj}=-{\cal I}_{jki}$, we have
\bea
{\cal T}[1,2,\cdots,n]&=&(-)^{n-2}{\cal T}[n,1]\cdot{\cal I}_{n21}\cdot{\cal I}_{n32}\cdots{\cal I}_{n(n-1)(n-2)}\nn
&=&(-)^n{\cal T}[n,n-1,\cdots,1]\,,
\eea
which immediately gives rise to \eref{color-reverse}.

The second relation is the generalized photon decoupling relation. Using the definition of the insertion operators, one
can decompose ${\cal I}_{1hn}$ as
\bea
{\cal I}_{1hn}={\cal I}_{1h2}+{\cal I}_{2h3}+\cdots+{\cal I}_{(n-1)hn}\,,
\eea
therefore
\bea
{\cal I}_{1h2}+{\cal I}_{2h3}+\cdots+{\cal I}_{(n-1)hn}+{\cal I}_{1hn}=0\,.
\eea
This algebraic relation indicates that
\bea
\sum_{\shuffle}\,{\cal A}(1,h\shuffle\{2,\cdots,n\})=0\,.
\eea
The shuffle of two ordered sets $\a\shuffle\b$ is the permutation of the set $\a\cup\b$ while preserving the ordering of $\a$ and $\b$.
For instance,
\bea
\sum_{\shuffle}\,\{1,2\}\shuffle\{3,4\}&=&\{1,2,3,4\}+\{1,3,2,4\}+\{1,3,4,2\}\nn
& &+\{3,1,2,4\}+\{3,1,4,2\}+\{3,4,1,2\}\,.
\eea

The third relation is the generalized Kleiss-Kuijf (KK) relation
\bea
{\cal A}(1,\alpha,n,\beta)=\sum_{\shuffle}\,(-)^{|\beta|}{\cal A}(1,\alpha\shuffle\beta^T,n)\,,~~~~\label{KK}
\eea
where $\a$ and $\b$ are two ordered sets, and $\b^T$ is obtained from $\b$ by reversing the ordering of elements.
To derive it, let us reformulate the trace operator ${\cal T}[a_1,\cdots,a_n]$ as
\bea
{\cal T}[a_1,\cdots,a_n]&=&{\cal T}[a_1,a_n]\cdot\prod_{i=2}^{n-1}{\cal I}_{a_{i-1}a_ia_n}\nn
&=&{\cal T}[a_1,a_n]\cdot\Big(\prod_{i=1}^k{\cal I}_{a_{i-1}a_ia_n}\Big)\cdot\Big((-)^{n-k-1}\prod_{j=k+1}^{n-1}{\cal I}_{a_na_ja_{j-1}}\Big)\,.
\eea
The operator ${\cal T}[a_1,a_n]\cdot\Big(\prod_{i=1}^k{\cal I}_{a_{i-1}a_ia_n}\Big)$ generates the color-ordering $(a_1,a_2,\cdots, a_k, a_n)$,
which is equivalent to $(a_n,a_1,\cdots, a_k)$, due to the cyclic symmetry. The operator $\Big((-)^{n-k-1}\prod_{j=k+1}^{n-1}{\cal I}_{a_na_ja_{j-1}}\Big)$ can be interpreted as inserting $\{a_{n-1},a_{n-2},\cdots,a_{k+1}\}$ between $a_n$ and $a_k$. More explicitly,
using the definition of the insertion operator we know that
\bea
{\cal I}_{ikj}={\cal I}_{ikl}+{\cal I}_{lkj}\,.
\eea
Repeating this decomposition, we find that the operator ${\cal I}_{a_na_{k+1}a_k}$
gives rise to
\bea
\sum_{\shuffle}\,{\cal A}(a_n,\{a_1,\cdots,a_{k-1}\}\shuffle a_{k+1},a_k)\,.
\eea
Applying the similar procedure recursively, it is easy to see that the operator $\Big((-)^{n-k-1}\prod_{j=k+1}^{n-1}{\cal I}_{a_na_ja_{j-1}}\Big)$ leads to
\bea
\sum_{\shuffle}\,{\cal A}(a_n,\{a_1,\cdots,a_{k-1}\}\shuffle \{a_{n-1},\cdots,a_{k+1}\},a_k)\,.
\eea
Using the above conclusion, and setting
$a_n=1$, $a_k=n$, $\{a_1,\cdots,a_{k-1}\}=\a$, $\{a_{k+1},\cdots,a_{n-1}\}=\b$, the KK relation \eref{KK} can be observed directly.

%%%%%%%%%%%%%%%%%%%%%%%%%%%%%%%%%%%%%%
\section{Discussion}
\label{secconclu}
%%%%%%%%%%%%%%%%%

With modifying the definition of the longitudinal operators, we have generalized various relations for on-shell amplitudes to
off-shell ones. The CHY formulism serve as a powerful tool for this work. As explained in \S\ref{secintro}, our result also provides a verification for the double-cover construction.

At the end of the paper, we want to point out three things.

First, the expansions of on-shell amplitudes, which can be derived via relations in Table \ref{tab:unifying}, can not be generalized to the off-shell case, since in the derivation proposed in \cite{Feng:2019cbe,Hu:2019qdq,Zhou:2019mbe}, the gauge invariance is a necessary tool. As mentioned in \S\ref{secreview}, for the off-shell case, the gauge invariance is lost. For some special case, for example the GR amplitude including two massive external legs with the same mass (a special case of the off-shell external massless states), one can avoid the appearing of corrections $\Delta_{ij}$ and $\eta_{ij}$ by choosing the removed rows and columns in the reduced matrix to be two massive ones, and obtain the reduced matrix totally the same as that for the massless on-shell amplitude. Then one can claim that the expansion of on-shell massless GR amplitude to BAS amplitudes also hold for this massive (or off-shell) GR amplitude, since from the CHY point of view, this expansion is just expanding ${\bf Pf}'\Psi^\Lambda$ to Parke-Taylor factors ${\cal C}^\Lambda_n(\sigma)$. But this manipulation can not be generalized to general off-shell amplitudes.

Secondly, the differential operators discussed in this paper will not affect any $k_i^2$. This fact has its physical meaning when treating amplitudes
with massive external states as off-shell CHY integrals. Thus, for the massive amplitudes of different theories related by differential
operators, the mass for a particular external state is unique, due to the invariance of $k_i^2$. For example, suppose the leg $1$ is a vector particle in
the amplitude ${\cal A}_1$, and is a scalar particle in the amplitude ${\cal A}_2$. If ${\cal A}_1$ and ${\cal A}_2$ are related by a
differential operator, then the vector particle and the scalar particle denoted by $1$ have the same mass.

Thirdly, in the literature \cite{Cheung:2017ems}, the gauge invariance is one of conditions for constructing differential operators. In the off-shell
case, the gauge invariance is lost. However, differential operators transmute off-shell amplitudes in the same manner as transmuting on-shell ones. It implies that maybe it is not suitable to regard the gauge invariance as the fundamental principle for constructing differential operators. The underlying principle which determines differential operators discussed in this paper is an interesting question.

%%%%%%%%%%%%%%%%%%%%%%%%%%%%%%%%%%%%%%%%%%%%%%%%%%
\section*{Acknowledgments}

This work is supported by Chinese NSF funding under
contracts No.11805163, as well as NSF of Jiangsu Province under Grant No.BK20180897.


\begin{thebibliography}{}


%%%%%%%%%%%%%%%%%%%%%%%%%%%%%%%%%%

%\cite{Cachazo:2013gna}
\bibitem{Cachazo:2013gna}
  F.~Cachazo, S.~He and E.~Y.~Yuan,
  ``Scattering equations and Kawai-Lewellen-Tye orthogonality,''
  Phys.\ Rev.\ D {\bf 90}, no. 6, 065001 (2014)
  doi:10.1103/PhysRevD.90.065001
  [arXiv:1306.6575 [hep-th]].
  %%CITATION = doi:10.1103/PhysRevD.90.065001;%%
  %113 citations counted in INSPIRE as of 16 Nov 2016

%\cite{Cachazo:2013hca}
\bibitem{Cachazo:2013hca}
  F.~Cachazo, S.~He and E.~Y.~Yuan,
  ``Scattering of Massless Particles in Arbitrary Dimensions,''
  Phys.\ Rev.\ Lett.\  {\bf 113}, no. 17, 171601 (2014)
  doi:10.1103/PhysRevLett.113.171601
  [arXiv:1307.2199 [hep-th]].
  %%CITATION = doi:10.1103/PhysRevLett.113.171601;%%
  %166 citations counted in INSPIRE as of 16 Nov 2016

%\cite{Cachazo:2013iea}
\bibitem{Cachazo:2013iea}
  F.~Cachazo, S.~He and E.~Y.~Yuan,
  ``Scattering of Massless Particles: Scalars, Gluons and Gravitons,''
  JHEP {\bf 1407}, 033 (2014)
  doi:10.1007/JHEP07(2014)033
  [arXiv:1309.0885 [hep-th]].
  %%CITATION = doi:10.1007/JHEP07(2014)033;%%
  %148 citations counted in INSPIRE as of 16 Nov 2016

%\cite{Cachazo:2014nsa}
\bibitem{Cachazo:2014nsa}
  F.~Cachazo, S.~He and E.~Y.~Yuan,
  ``Einstein-Yang-Mills Scattering Amplitudes From Scattering Equations,''
  JHEP {\bf 1501}, 121 (2015)
  doi:10.1007/JHEP01(2015)121
  [arXiv:1409.8256 [hep-th]].
  %%CITATION = doi:10.1007/JHEP01(2015)121;%%
  %64 citations counted in INSPIRE as of 16 Nov 2016

%\cite{Cachazo:2014xea}
\bibitem{Cachazo:2014xea}
  F.~Cachazo, S.~He and E.~Y.~Yuan,
  ``Scattering Equations and Matrices: From Einstein To Yang-Mills, DBI and NLSM,''
  JHEP {\bf 1507}, 149 (2015)
  doi:10.1007/JHEP07(2015)149
  [arXiv:1412.3479 [hep-th]].
  %%CITATION = doi:10.1007/JHEP07(2015)149;%%
  %82 citations counted in INSPIRE as of 16 Nov 2016

%%%%%%%%%%%%%%%%%%%%%%%%%%%%%%%%%%%%%%%%%%%%%

%\cite{Cheung:2017ems}
\bibitem{Cheung:2017ems}
  C.~Cheung, C.~H.~Shen and C.~Wen,
  ``Unifying Relations for Scattering Amplitudes,''
  JHEP {\bf 1802}, 095 (2018)
  doi:10.1007/JHEP02(2018)095
  [arXiv:1705.03025 [hep-th]].
  %%CITATION = doi:10.1007/JHEP02(2018)095;%%
  %22 citations counted in INSPIRE as of 12 Aug 2018

%%%%%%%%%%%%%%%%%%%%%%%%%%%%%%%%%%%%%%%%%%%%%%

%\cite{Zhou:2018wvn,Bollmann:2018edb}
\bibitem{Zhou:2018wvn}
  K.~Zhou and B.~Feng,
  ``Note on differential operators, CHY integrands, and unifying relations for amplitudes,''
  JHEP {\bf 1809}, 160 (2018)
 % doi:10.1007/JHEP09(2018)160
  [arXiv:1808.06835 [hep-th]].
  %%CITATION = doi:10.1007/JHEP09(2018)160;%%
  %3 citations counted in INSPIRE as of 18 Mar 2019

%\cite{Bollmann:2018edb}
\bibitem{Bollmann:2018edb}
  M.~Bollmann and L.~Ferro,
  ``Transmuting CHY formulae,''
  JHEP {\bf 1901}, 180 (2019)
  %doi:10.1007/JHEP01(2019)180
  [arXiv:1808.07451 [hep-th]].
  %%CITATION = doi:10.1007/JHEP01(2019)180;%%
  %2 citations counted in INSPIRE as of 18 Mar 2019

%%%%%%%%%%%%%%%%%%%%%%%%%%%%%%%%%%%%%%%%%%%%%

%\cite{Feng:2019cbeZhou:2019mbe}
\bibitem{Feng:2019cbe}
  B.~Feng, X.~Li and K.~Zhou,
  ``Expansion of Einstein-Yang-Mills theory by differential operators,''
  Phys.\ Rev.\ D {\bf 100}, no. 12, 125012 (2019)
  doi:10.1103/PhysRevD.100.125012
  [arXiv:1904.05997 [hep-th]].
  %%CITATION = doi:10.1103/PhysRevD.100.125012;%%
  %4 citations counted in INSPIRE as of 20 Mar 2020

%\cite{Hu:2019qdq}
\bibitem{Hu:2019qdq}
  S.~Q.~Hu and K.~Zhou,
  ``Expansion of tree amplitudes for EM and other theories,''
  arXiv:1907.07857 [hep-th].
  %%CITATION = ARXIV:1907.07857;%%

%\cite{Zhou:2019mbe}
\bibitem{Zhou:2019mbe}
  K.~Zhou,
  ``Unified web for expansions of amplitudes,''
  JHEP {\bf 1910}, 195 (2019)
  doi:10.1007/JHEP10(2019)195
  [arXiv:1908.10272 [hep-th]].
  %%CITATION = doi:10.1007/JHEP10(2019)195;%%
  %2 citations counted in INSPIRE as of 20 Mar 2020


%%%%%%%%%%%%%%%%%%%%%%%%%%%%%%%%%%%%%%%%%%%%%%%%
%\cite{Stieberger:2016lng,Schlotterer:2016cxa,Chiodaroli:2017ngp,DelDuca:1999rs,Nandan:2016pya,delaCruz:2016gnm,Fu:2017uzt,Teng:2017tbo,Du:2017kpo,Du:2017gnh}
\bibitem{Stieberger:2016lng}
  S.~Stieberger and T.~R.~Taylor,
  ``New relations for Einstein-Yang-Mills amplitudes,''
  Nucl.\ Phys.\ B {\bf 913}, 151 (2016)
 % doi:10.1016/j.nuclphysb.2016.09.014
  [arXiv:1606.09616 [hep-th]].
  %%CITATION = doi:10.1016/j.nuclphysb.2016.09.014;%%
  %39 citations counted in INSPIRE as of 18 Mar 2019


%\cite{Schlotterer:2016cxa}
\bibitem{Schlotterer:2016cxa}
  O.~Schlotterer,
  ``Amplitude relations in heterotic string theory and Einstein-Yang-Mills,''
  JHEP {\bf 1611}, 074 (2016)
  %doi:10.1007/JHEP11(2016)074
  [arXiv:1608.00130 [hep-th]].
  %%CITATION = doi:10.1007/JHEP11(2016)074;%%
  %23 citations counted in INSPIRE as of 18 Mar 2019


%\cite{Chiodaroli:2017ngp}
\bibitem{Chiodaroli:2017ngp}
  M.~Chiodaroli, M.~Gunaydin, H.~Johansson and R.~Roiban,
  ``Explicit Formulae for Yang-Mills-Einstein Amplitudes from the Double Copy,''
  JHEP {\bf 1707}, 002 (2017)
  doi:10.1007/JHEP07(2017)002
  [arXiv:1703.00421 [hep-th]].
  %%CITATION = doi:10.1007/JHEP07(2017)002;%%
  %35 citations counted in INSPIRE as of 31 May 2019

%\cite{DelDuca:1999rs}
\bibitem{DelDuca:1999rs}
  V.~Del Duca, L.~J.~Dixon and F.~Maltoni,
  ``New color decompositions for gauge amplitudes at tree and loop level,''
  Nucl.\ Phys.\ B {\bf 571}, 51 (2000)
  doi:10.1016/S0550-3213(99)00809-3
  [hep-ph/9910563].
  %%CITATION = doi:10.1016/S0550-3213(99)00809-3;%%
  %225 citations counted in INSPIRE as of 06 Jun 2019


%\cite{Nandan:2016pya}
\bibitem{Nandan:2016pya}
  D.~Nandan, J.~Plefka, O.~Schlotterer and C.~Wen,
  ``Einstein-Yang-Mills from pure Yang-Mills amplitudes,''
  JHEP {\bf 1610}, 070 (2016)
  %doi:10.1007/JHEP10(2016)070
  [arXiv:1607.05701 [hep-th]].
  %%CITATION = doi:10.1007/JHEP10(2016)070;%%
  %34 citations counted in INSPIRE as of 18 Mar 2019


%\cite{delaCruz:2016gnm}
\bibitem{delaCruz:2016gnm}
  L.~de la Cruz, A.~Kniss and S.~Weinzierl,
  ``Relations for Einstein-Yang-Mills amplitudes from the CHY representation,''
  Phys.\ Lett.\ B {\bf 767}, 86 (2017)
 % doi:10.1016/j.physletb.2017.01.036
  [arXiv:1607.06036 [hep-th]].
  %%CITATION = doi:10.1016/j.physletb.2017.01.036;%%
  %25 citations counted in INSPIRE as of 18 Mar 2019

%\cite{Fu:2017uzt}
\bibitem{Fu:2017uzt}
  C.~H.~Fu, Y.~J.~Du, R.~Huang and B.~Feng,
  ``Expansion of Einstein-Yang-Mills Amplitude,''
  JHEP {\bf 1709}, 021 (2017)
  %doi:10.1007/JHEP09(2017)021
  [arXiv:1702.08158 [hep-th]].
  %%CITATION = doi:10.1007/JHEP09(2017)021;%%
  %14 citations counted in INSPIRE as of 18 Mar 2019


%\cite{Teng:2017tbo}
\bibitem{Teng:2017tbo}
  F.~Teng and B.~Feng,
  ``Expanding Einstein-Yang-Mills by Yang-Mills in CHY frame,''
  JHEP {\bf 1705}, 075 (2017)
  %doi:10.1007/JHEP05(2017)075
  [arXiv:1703.01269 [hep-th]].
  %%CITATION = doi:10.1007/JHEP05(2017)075;%%
  %19 citations counted in INSPIRE as of 18 Mar 2019

%\cite{Du:2017kpo}
\bibitem{Du:2017kpo}
  Y.~J.~Du and F.~Teng,
  ``BCJ numerators from reduced Pfaffian,''
  JHEP {\bf 1704}, 033 (2017)
 % doi:10.1007/JHEP04(2017)033
  [arXiv:1703.05717 [hep-th]].
  %%CITATION = doi:10.1007/JHEP04(2017)033;%%
  %19 citations counted in INSPIRE as of 18 Mar 2019


%\cite{Du:2017gnh}
\bibitem{Du:2017gnh}
  Y.~J.~Du, B.~Feng and F.~Teng,
  ``Expansion of All Multitrace Tree Level EYM Amplitudes,''
  JHEP {\bf 1712}, 038 (2017)
 % doi:10.1007/JHEP12(2017)038
  [arXiv:1708.04514 [hep-th]].
  %%CITATION = doi:10.1007/JHEP12(2017)038;%%
  %16 citations counted in INSPIRE as of 18 Mar 2019

 %%%%%%%%%%%%%%%%%%%%%%%%%%%%%%%%%%%%%%%%%%%%%%%%%%%%



%\cite{Gomez:2016bmv}
\bibitem{Gomez:2016bmv}
  H.~Gomez,
  ``$\Lambda$ scattering equations,''
  JHEP {\bf 1606}, 101 (2016)
  doi:10.1007/JHEP06(2016)101
  [arXiv:1604.05373 [hep-th]].
  %%CITATION = doi:10.1007/JHEP06(2016)101;%%
  %34 citations counted in INSPIRE as of 19 Feb 2020

%\cite{Cardona:2016bpi}
\bibitem{Cardona:2016bpi}
  C.~Cardona and H.~Gomez,
  ``Elliptic scattering equations,''
  JHEP {\bf 1606}, 094 (2016)
  doi:10.1007/JHEP06(2016)094
  [arXiv:1605.01446 [hep-th]].
  %%CITATION = doi:10.1007/JHEP06(2016)094;%%
  %31 citations counted in INSPIRE as of 13 Mar 2020

%\cite{Bjerrum-Bohr:2018lpz}
\bibitem{Bjerrum-Bohr:2018lpz}
  N.~E.~J.~Bjerrum-Bohr, P.~H.~Damgaard and H.~Gomez,
  ``New Factorization Relations for Yang Mills Amplitudes,''
  Phys.\ Rev.\ D {\bf 99}, no. 2, 025014 (2019)
  doi:10.1103/PhysRevD.99.025014
  [arXiv:1810.05023 [hep-th]].
  %%CITATION = doi:10.1103/PhysRevD.99.025014;%%
  %5 citations counted in INSPIRE as of 17 Feb 2020

%\cite{Gomez:2018cqg}
\bibitem{Gomez:2018cqg}
  H.~Gomez,
  ``Scattering equations and a new factorization for amplitudes. Part I. Gauge theories,''
  JHEP {\bf 1905}, 128 (2019)
  doi:10.1007/JHEP05(2019)128
  [arXiv:1810.05407 [hep-th]].
  %%CITATION = doi:10.1007/JHEP05(2019)128;%%
  %5 citations counted in INSPIRE as of 19 Feb 2020

%\cite{Bjerrum-Bohr:2018jqe}
\bibitem{Bjerrum-Bohr:2018jqe}
  N.~E.~J.~Bjerrum-Bohr, H.~Gomez and A.~Helset,
  ``New factorization relations for nonlinear sigma model amplitudes,''
  Phys.\ Rev.\ D {\bf 99}, no. 4, 045009 (2019)
  doi:10.1103/PhysRevD.99.045009
  [arXiv:1811.06024 [hep-th]].
  %%CITATION = doi:10.1103/PhysRevD.99.045009;%%
  %6 citations counted in INSPIRE as of 19 Feb 2020

%\cite{Gomez:2019cik}
\bibitem{Gomez:2019cik}
  H.~Gomez and A.~Helset,
  ``Scattering equations and a new factorization for amplitudes. Part II. Effective field theories,''
  JHEP {\bf 1905}, 129 (2019)
  doi:10.1007/JHEP05(2019)129
  [arXiv:1902.02633 [hep-th]].
  %%CITATION = doi:10.1007/JHEP05(2019)129;%%
  %4 citations counted in INSPIRE as of 13 Mar 2020

%%%%%%%%%%%%%%%%%%%%%%%%%%%%%%%%%%%%%%%%%%%%%%%%%%%%%%%


%\cite{Naculich:2015zha,Lam:2019mfk,Bjerrum-Bohr:2019nws}
\bibitem{Naculich:2015zha}
S.~G.~Naculich,
``CHY representations for gauge theory and gravity amplitudes with up to three massive particles,''
JHEP \textbf{05}, 050 (2015)
doi:10.1007/JHEP05(2015)050
[arXiv:1501.03500 [hep-th]].
%40 citations counted in INSPIRE as of 25 May 2020

%\cite{Lam:2019mfk,Bjerrum-Bohr:2019nws}
\bibitem{Lam:2019mfk}
C.~Lam,
``Off-shell Yang-Mills amplitude in the Cachazo-He-Yuan formalism,''
Phys. Rev. D \textbf{100}, no.4, 045009 (2019)
doi:10.1103/PhysRevD.100.045009
[arXiv:1905.05101 [hep-th]].
%3 citations counted in INSPIRE as of 25 May 2020

%\cite{Bjerrum-Bohr:2019nws}
\bibitem{Bjerrum-Bohr:2019nws}
N.~Bjerrum-Bohr, A.~Cristofoli, P.~H.~Damgaard and H.~Gomez,
``Scalar-Graviton Amplitudes,''
JHEP \textbf{11}, 148 (2019)
doi:10.1007/JHEP11(2019)148
[arXiv:1908.09755 [hep-th]].
%6 citations counted in INSPIRE as of 25 May 2020



\end{thebibliography}
\end{document}